\documentclass[fleqn,10pt]{wlscirep}
\usepackage[utf8]{inputenc}
\usepackage[T1]{fontenc}
\usepackage{lineno}

\usepackage{booktabs}
\usepackage{multirow}
\usepackage{longtable}
\usepackage{xr-hyper}
\usepackage{xcolor}
\usepackage{hyperref}

\hypersetup{
pdftitle={Mapping urban segregation through co-residence network reconstruction},
pdfsubject={Manuscript},
pdfkeywords={Vienna Mosaic, Social Borders, Melting Pot}
}

\makeatletter
\newcommand*{\addFileDependency}[1]{
  \typeout{(#1)}
  \@addtofilelist{#1}
  \IfFileExists{#1}{}{\typeout{No file #1.}}
}
\makeatother

\title{Mapping urban segregation through co-residence network reconstruction}

\author[1,2,*]{Marc Sadurní}
\author[5,3,4]{Samuel Martin-Gutierrez}
\author[3]{Ola Ali}
\author[3,4]{Ana María Jaramillo}
\author[3]{Rafael Prieto-Curiel}
\author[4,3,*]{Fariba Karimi}
\affil[1]{Departament de Física de la Matèria Condensada, Universitat de Barcelona, 08028 Barcelona, Catalonia, Spain}
\affil[2]{Universitat de Barcelona Institute of Complex Systems, 08028 Barcelona, Catalonia, Spain}
\affil[3]{Complexity Science Hub, 1030 Vienna, Austria}
\affil[4]{Graz University of Technology, 8010 Graz, Austria}
\affil[5]{Grupo de Sistemas Complejos, Universidad Politécnica de Madrid, 28040 Madrid, Spain}
\affil[*]{Corresponding authors: Marc Sadurní (E-mail: marc.sadurni@ub.edu); Fariba Karimi (E-mail: karimi@csh.ac.at)}

\begin{abstract} \label{sec:abstract}
Urban segregation poses a critical challenge for cities, exacerbating inequalities, social tensions, fears, and polarisation. It emerges from the interplay of socio-economic disparities, housing constraints, and residential preferences, and can disproportionately affect migrant communities. Here, we study residential segregation in Vienna using a city-wide administrative snapshot of registered residents, covering the full foreign population and Austrian nationals at the district level. We introduce a network-based approach by constructing a statistically validated co-residence network in which nodes represent nationalities and links capture whether pairs of groups live in the same districts more or less often than expected under a population-size-preserving null model. Applying community detection to this network reveals two major clusters of nationalities with distinct co-residence patterns. These clusters are systematically associated with district-level income disparities and diversity, while also reflecting the geographical proximity of countries of origin, with nationalities from nearby regions tending to share similar residential patterns within Vienna. Our results show how network methods can provide an intuitive and interpretable map of urban residential sorting, complementing traditional segregation indices and highlighting the multiple dimensions underlying migrant integration in diverse cities.
\end{abstract}
\begin{document}

\flushbottom
\maketitle
\section*{Introduction} \label{sec:introduction}
{
Cities offer unparalleled access to essential services, including healthcare, education, transportation, cultural amenities, water, and sanitation services \cite{Glaeser01,Leon08,Winters11}. Cities are hubs of social interaction, fostering productivity and attracting talent \cite{keuschnigg2019urban}. Some of those positive urban features are further accentuated in large cities. For example, large cities tend to attract more people, primarily young people \cite{PrietoColombiaMobility}, and have more diversity and a much larger share of international migrants \cite{ScalingMigrationRPC}. Although large cities could be ideal for social mixing, they can also form highly unequal and segregated communities, where groups rarely interact with others \cite{sousa2022quantifying}. In some cities, for example, wealthy neighbourhoods tend to segregate themselves from the urban population and create private areas \cite{ParnellAfrica}, often motivated by fear and prejudice \cite{caldeira2000city}. Moreover, although residential segregation can strengthen ethnic ties and support newcomers, it constrains social ties with natives in terms of support, advice, and frequency of contact \cite{vervoort2012ethnic}, resulting in reduced access to essential services \cite{dinwiddie2013residential}. Segregated cities often result from the residential selection process, in which even small preferences for certain neighbourhood characteristics can significantly alter the distribution of population groups \cite{clark2008understanding,bolt2008minority,lee2016temporal}.
}

{
The emergent phenomenon of segregation significantly affects the design and planning of cities \cite{bruch2019choice}. In principle, most cities are not designed to be segregated, but individual residential choices, governmental programs \cite{trounstine2018segregation}, mobility patterns \cite{candipan2021residence}, and preferences for living near similar others may result in segregated cities \cite{clark2008understanding}. Fuelled by trust, mutual understanding, shared values, and behaviours, the social capital arising from segregation aims to preserve culture for immigrant and local communities and to maintain socio-economic positions \cite{portes1993new,light2019segregation,ParnellAfrica}. Thus, segregation can also be a mechanism by which communities create and preserve social capital \cite{Bourdieu1986}. However, high levels of economic, social, and geographical segregation accentuate social, health, and education inequalities \cite{DEVERTEUIL2009433,sadurni2025emergence}. Underprivileged communities are at risk of isolation, marginalization, and victimization \cite{o2007crime,desmond2015forced}, and intergroup conflict can emerge from hostility and cultural misunderstandings \cite{pettigrew2006meta,helbing2015saving}. The interplay among these dimensions of segregation necessitates that global cities avoid becoming disintegrated ghettos and address the effects of segregation on increasing polarisation, political extremism, and perceptions of minorities \cite{BoltzmannSegregation,motyl2014ideological,lee2019homophily}.
}

{
Vienna is the capital and largest city of Austria, with approximately 2 million inhabitants across its 23 districts \cite{viennapopulationtotal}. Its central position makes the city a key migration hub in Europe and a unique case for studying segregation patterns, particularly given its historical context and its status as a leading city for quality of life \cite{GlobLiveIndex}, achieved through policies promoting affordable housing and social integration. In 2023, nearly 40\% of Austria's foreign residents lived in Vienna \cite{viennapopulationtotal}. This is especially true for refugees, who often choose to resettle in Vienna due to the city’s support systems and services for asylum seekers \cite{stability_refugee}. Vienna also stands out for its extensive history of social housing, with one of the largest stocks of community-owned buildings in Europe \cite{Housing_Integration_Vienna,socialhousingvienna}. This combination of a high foreign population and a strong tradition of public housing creates a relevant setting in which residential segregation and district-level sorting can be explored in depth. Previous studies have documented spatially segregated patterns of migration in Vienna \cite{Housing_Integration_Vienna,kohlbacher2020globalization}, and established segregation indices quantify important dimensions of unevenness and diversity. Here, we focus on a different but related question: how nationalities are relationally organised through shared district-level residence patterns.
}

{
A central challenge in segregation research is to move from group-specific measures of unevenness to the broader structure that emerges among population groups across urban space. Established indices such as dissimilarity or diversity measures are well suited to quantifying how unevenly individual groups are distributed across spatial units. However, they do not directly show which groups share similar residential environments, or how these pairwise associations organise into larger structures. Network analysis provides a complementary perspective: by representing nationalities as nodes and shared residence patterns as links, it becomes possible to study segregation as a relational structure among groups, rather than only as a set of group-specific summary values. This perspective makes it possible to identify which groups tend to be registered in the same districts, how these pairwise associations form larger clusters, and how such clusters relate to district-level socio-economic and demographic characteristics. Community detection then offers an interpretable way to identify mesoscopic residential groupings and examine how they relate to district income, diversity, and the geographical proximity of countries of origin.
}

{
To investigate these dynamics, we use administrative residence-registration data from Vienna, based on the mandatory registration form, the ``Meldezettel''. These data provide a city-wide snapshot of registered residence locations, aggregated at the district level. We develop a statistical measure of district-level shared residence patterns among nationalities. Importantly, we do not interpret segregation as the presence of mono-national enclaves, but rather as uneven exposure and differential concentration of nationality groups across the city. First, we identify pairs of groups that are registered in the same districts more or less often than expected under a population-size-preserving null model. We then construct a network linking nationalities through significant shared residence patterns and apply community detection to identify large-scale residential groupings. This approach reveals two prominent clusters with distinct district-level residence patterns. These clusters are systematically associated with district-level income and diversity, and their structure also reflects the geographical proximity of countries of origin, with nationalities from nearby regions tending to share similar residence patterns within Vienna. The resulting framework complements traditional segregation indices by providing an interpretable map of how national groups are relationally organised across the city. However, given the level of aggregation, these patterns cannot be directly interpreted as evidence of fine-scale spatial proximity, interpersonal contact, or individual residential preferences.
}

\section*{Methods}\label{sec:methods}

\begin{figure*}[bt!]
    \centering
    \includegraphics[width=\textwidth]{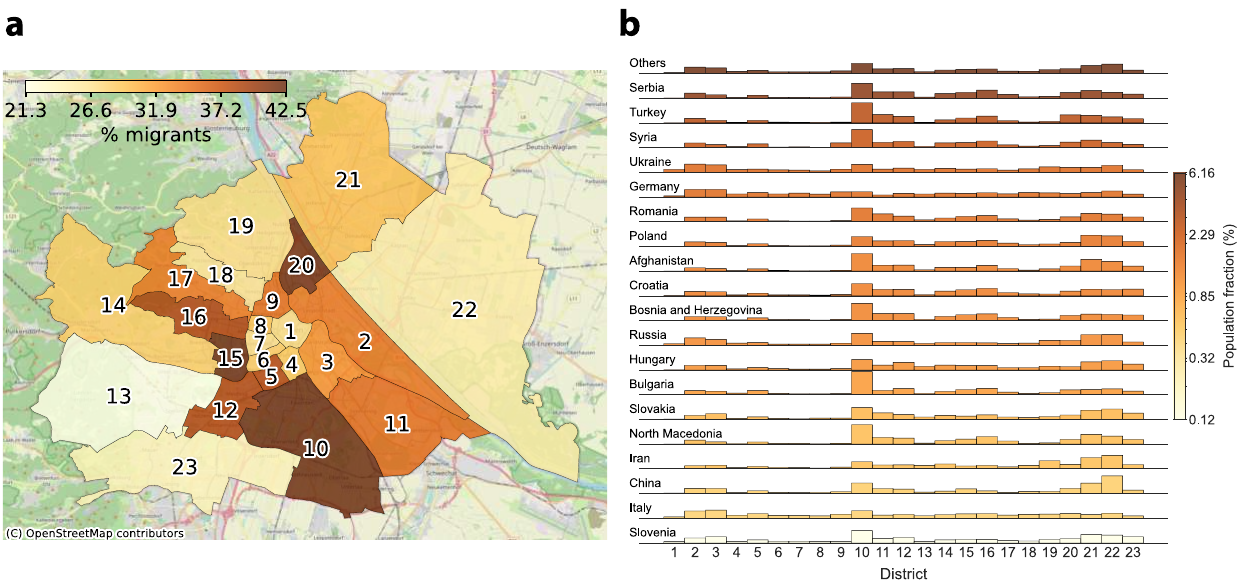}
    \caption{\textbf{Melting Pot in Vienna.} \textbf{a} Population of foreigners in Vienna in 2023 (in \%). Districts 15 and 13 have the highest and lowest percentage of foreigners, respectively. \textbf{b} Population distributions of the top 19 most populous foreign nationalities in Vienna. The colour bar indicates the fraction of Vienna's population on a log scale. Countries are sorted from the highest to the lowest population fraction.}
\label{fig:fig1}
\end{figure*}

{
\subsection*{Residence registration data of foreigners in Vienna} 
The data used here is provided by the Federal Ministry of Interior of Austria, Bundesministerium für Inneres ``BMI''. The data includes all individuals in Austria with a foreign nationality who register their residence using the mandatory registration form (the ``Meldezettel''). The form is updated every time a person changes their primary residence. The dataset thus encompasses all foreigners residing in Vienna up until September 2023. The data excludes short-term visitors, such as tourists who are not required to register, and does not account for undocumented foreigners (Supplementary Note \ref{sec:residencedata}). For this study, we used a snapshot of the residence locations of the entire foreign population in Vienna, aggregated at the district level and captured on 22nd September 2023 (Fig. \ref{fig:fig1}\textbf{a}). Our objective is to examine district-level residential sorting across nationalities using administrative residence registrations. As an administrative scale, districts are best suited to capturing broad residential sorting rather than fine-scale neighbourhood proximity or direct interpersonal exposure. However, in Vienna they also constitute meaningful urban units through which residents encounter differences in housing markets, public services, infrastructure, and the broader sociodemographic composition of their residential environment. While smaller spatial units, such as census tracts \cite{KlaiberMorawetz2021WelfareImpacts}, can capture more granular forms of segregation, the district level provides a consistent and interpretable city-wide scale for comparing the residential distributions of national groups. In the Supplementary Notes, we also examine experienced segregation by accounting for mobility to adjacent districts within the city.
}

{
\paragraph{Residence district locations' extraction.} To protect individual privacy, population counts are aggregated at the district level, with statistics provided only for nationalities represented by at least ten individuals within a district. By setting this threshold, we mitigate the risk of identifying specific individuals within each district, which is particularly important in small population groups or areas with fewer residents of a given nationality. Aggregating data at this level helps maintain confidentiality by concealing individual characteristics within broader statistical averages. This approach ensures that privacy standards are upheld while still allowing for meaningful insights into district-level patterns and trends in foreign residency. This method identifies the top 19 most populous foreign nationalities in Vienna and consolidates the remaining nationalities under the category ``\textit{Others}'' (Fig. \ref{fig:fig1}\textbf{b}). By focusing on these top 19 countries, we capture the most significant demographic groups contributing to Vienna's foreign population while ensuring data manageability and compliance with privacy regulations. The ``\textit{Others}'' category aggregates the less-represented nationalities, allowing us to maintain a comprehensive coverage of all groups in the analysis without risking the identification of individuals from smaller communities.
}

{
\subsection*{Complementary datasets} To support our analyses of the residential patterns, we use additional datasets. The geographic boundaries for the Vienna districts were obtained from the Vienna GIS database \cite{viennadistricts}, while the global GIS data provided country boundaries to facilitate spatial visualisation and cross-national comparisons \cite{countriesGis}. Population counts by district were sourced from census data in 2023 \cite{viennapopulation}. The average net income at the district level in 2020 served as a proxy for economic status, helping to assess the socio-economic factors that influence residential clustering \cite{viennaincome}. Additional datasets are detailed in the Supplementary Note \ref{sec:otherdatasets}.
}

\subsection*{Mapping district-level shared residence patterns}

{
To explore district-level residential sorting, we analyse the rates at which individuals from different nationalities reside in the same districts. First, we map pairwise associations between countries using district-level residence counts. Next, using statistical tests, we estimate the significance of each association, thereby removing random associations that can be explained by differences in district size and nationality population size. This procedure yields a network in which links indicate statistically significant shared residence at the district level. The links between nationalities are based on the degree to which two nationalities are overrepresented or underrepresented in the same districts relative to a null model. We adapt a methodology originally applied to study common information interests across countries  \cite{karimi2015mapping}. To focus on cross-national shared residence patterns, self-loops representing within-nationality concentration are excluded from this network. Finally, we use a network community-detection method based on random walks to identify groups of countries from pairwise connections. This approach provides an interpretable map of the large-scale structure of district-level residential sorting in Vienna \cite{karimi2015mapping}.
}

{
\paragraph{Extracting shared residence links.} Our goal is to establish connections between countries based on the district-level overlap of their registered residents. We first aggregate the total number of individuals from each nationality residing in each district of Vienna (Fig. \ref{fig:fig1}\textbf{b}). To quantify shared residence at the district level, we use a well-mixed district-level baseline and ask how likely it is that two individuals randomly sampled from the same district belong to two specific nationalities. This provides a simple measure of the overlap between national groups within each district. For a given district $d$, if the population of country $i$ in that district is $\kappa_{i}^d$, the metric for district-level shared residence between countries $i$ and $j$ is given by:
\begin{equation}
    w_{ij}^d = \frac{1}{(N^d)^2}\kappa_{i}^d  \kappa_{j}^d \propto \kappa_{i}^d  \kappa_{j}^d,
\end{equation}
where $N^d$ is the total population of the district. In practice, we define the co-residence metric as $w_{ij}^d = \kappa_{i}^d  \kappa_{j}^d$ as the empirical district-level shared residence weight. Because this raw weight depends on both nationality population sizes and district populations, we next normalize it using a null model that controls for these size effects. This weight captures the extent to which two nationalities overlap within the same district, based on their aggregate residential distributions. Repeating this calculation for all pairs of nationalities yields a fully connected network in which nodes are countries and weighted links represent shared residence at the district level.
}

{
\paragraph{Extracting significant shared residence links.} Using raw shared residence weights introduces bias due to variation in nationality population sizes and district populations. This can produce strong links driven primarily by large groups or densely populated districts, rather than by statistically distinctive patterns of shared residence. To account for this, we compare the empirically observed link weights to a null model. We employ a statistical validation method that filters out connections likely resulting from size effects or random placement. Specifically, we randomise residence assignments while keeping the total populations of countries and districts fixed (Fig. \ref{fig:fig2}\textbf{a}). We then use the randomised resident counts as a baseline. If the empirical data show that individuals from two countries are registered in the same districts more frequently than expected under this baseline, we interpret this as a positive association between the two countries. Conversely, if their shared residence is lower than expected, we interpret this as a negative association. Formally, this method uses a multinomial distribution to provide analytical expressions for the average and standard deviation of shared residence weights, thereby identifying significant links in a bipartite system with countries on one side and districts on the other (Supplementary Fig. \ref{fig:figs5}) \cite{karimi2015mapping}. Other methods exist to evaluate significant correlations between entities in bipartite systems \cite{zweig2011systematic, tumminello2011statistically, serrano2009extracting}. Nonetheless, our model offers the advantage of readily accounting for variation in both district populations and nationality population sizes.
}

Under this null model, the nationality composition of each district is generated by drawing residents from the city-wide nationality distribution while keeping district populations fixed. This randomisation provides the baseline expectation for how often two nationalities would be registered in the same district by chance alone, given their overall presence in the city. From this null model, the analytical expected shared residence weight $\mu_{ij}^d$ for nationalities $i$ and $j$ in district $d$ can be expressed as \cite{karimi2015mapping}:
\begin{equation}
\mu_{i j}^d=N^d\left(N^d-1\right) P_i P_j,
\end{equation}
where $N^d$ is the total population in the district $d$, and $P_i=\sum_d \kappa_i^d / N_{\text {City}}$ represents the proportion of residents of nationality $i$ relative to the total population of Vienna, $N_{\text {City}}=\sum_d N^d$.

{
To evaluate the significance of shared residence patterns, we calculate a standardised \textit{z}-score for each nationality pair $(i, j)$ across districts. This score indicates how much the observed district-level overlap between the two nationalities deviates from the expected value under the null model. For a pair $(i, j)$ in district $d$, the \textit{z}-score is defined as:
\begin{equation}
z_{ij}^d= \frac{w_{i j}^d-\mu_{i j}^d}{\sigma_{i j}^d},
\end{equation}
where $\sigma_{ij}^d$ is the standard deviation of the null distribution for link weights. Using the multinomial theorem multiple times, one can compute the standard deviation as \cite{karimi2015mapping}:
\begin{equation}
\left(\sigma_{i j}^d\right)^2=N^d\left(N^d-1\right) P_i P_j\left(\left(6-4 N^d\right) P_i P_j+\left(N^d-2\right)\left(P_i+P_j\right)+1\right).
\end{equation}

The \textit{z}-score standardises observed shared residence weights, allowing comparisons between nationality pairs regardless of population sizes or district populations. Finally, to capture the overall district-level association between nationalities $i$ and $j$ across Vienna, we aggregate the $\textit{z}$-scores over all districts to compute the cumulative \textit{z}-score for each pair of countries $i$ and $j$ (Fig. \ref{fig:fig2}\textbf{a}):
\begin{equation}
z_{ij} = \sum_d z_{ij}^d.
\label{eq:zscore}
\end{equation}

These cumulative \textit{z}-scores allow us to compare shared residence weights between nationality pairs while accounting for variation across districts. Higher cumulative \textit{z}-scores for a pair $(i, j)$ indicate that the two nationalities are registered in the same districts more often than expected under the null model. Lower cumulative \textit{z}-scores indicate that the two nationalities are registered in the same districts less often than expected. Among the 21 groups (19 most populous foreign nationalities, \textit{Others}, and Austrians), we identified 210 significant links, forming a fully connected, undirected, and weighted network based on district-level shared residence (Fig. \ref{fig:fig2}\textbf{b}).

Positive links indicate that, on average, the district-level overlap between residents from a pair of countries exceeds the expected value. In contrast, negative links indicate that the observed overlap is lower than expected by chance. The network has 80 positive links (shown in red) and 130 negative links (in blue). See GitHub repository \cite{github} for a detailed list of $\textit{z}$-score values. This raises the question of what underlying factors drive these patterns of association. To address this question, we begin by identifying clusters of nationalities connected by positive shared residence links, filtering out the negative links and considering exclusively the 80 positive connections.
}

{
\paragraph{Clustering nationalities based on district co-residence patterns.} To identify large-scale residential patterns within Vienna, we cluster nationalities according to how similarly their residents are distributed across the city’s districts. By clustering these connections, we aim to uncover large-scale structures of district-level shared residence among national communities within the city.

To determine these clusters, we employ a community-detection approach that uses the strength of co-residence links among nationalities across Vienna's districts as the basis for grouping. The shared residence network is constructed with each nationality pair $(i,j)$ linked by their co-residence weights $z_{ij}$, which reflect the degree of district-level overlap between the two nationalities relative to the null model (Fig. \ref{fig:fig2}\textbf{b}). This network provides the structural foundation to reveal groups of countries with significant shared residence patterns. Our approach to clustering relies on the network community-detection method known as the map equation, which leverages random walks to detect communities based on their flow within the network \cite{rosvall2008maps,rosvall2009map}. In this context, the map equation algorithm simulates a sequence of ``visits'' to nationalities, where movement between nationalities is determined by the strength of their co-residence link. This method clusters nationalities based on the frequency with which a random walker transitions between them, prioritising strongly interconnected groups whose residents show similar district-level residence patterns. Nationalities with stronger internal connections are likely to retain them for extended periods, resulting in distinct, identifiable clusters. Here, we use the map equation’s associated search algorithm, Infomap \cite{mapequation2023software}, to identify groups of countries. The number of communities is not imposed a priori: Infomap returns an optimal partition of the country network that minimises the map-equation objective (i.e., the description length of a random walk on the weighted network).

For our data, Infomap uses the weights of links ($\textit{z}$-scores) once they are filtered using the Bonferroni correction (Supplementary Note \ref{sec:bonferrini}) in order to determine the clusters of strongly connected countries (Fig. \ref{fig:fig2}\textbf{c}). It consists of two clusters, except for China, Slovakia, Afghanistan, and Poland (as these countries become disconnected and are no longer assigned to any cluster). This disconnection reflects the fact that these nationalities exhibit distinct district-level residence patterns in Vienna, differing significantly from other countries in their shared residence links (Supplementary Fig. \ref{fig:figs4}).

To gain better insights into the shape of these two clusters, we examine the
country network structure within each cluster (Fig. \ref{fig:fig2}\textbf{d}). People with the same nationality may connect through shared background, and people living in the same district may interact due to spatial proximity. We performed additional robustness checks, including clustering results without Bonferroni correction for multiple comparisons and varying Infomap’s Markov time parameter, to assess the stability of the community structure (Supplementary Note \ref{sec:sanity}). 

The method captures not only direct pairwise connections but also implicit relationships in which two nationalities may show similar district-level residence patterns through common links with other nationalities, even if they are not strongly connected to each other directly. The resulting clusters thus provide a simplified yet meaningful representation of residential sorting among national communities, offering a macroscopic perspective on their spatial organisation within Vienna.

\begin{figure*}[bt!]
\centering
\includegraphics[width=\textwidth]{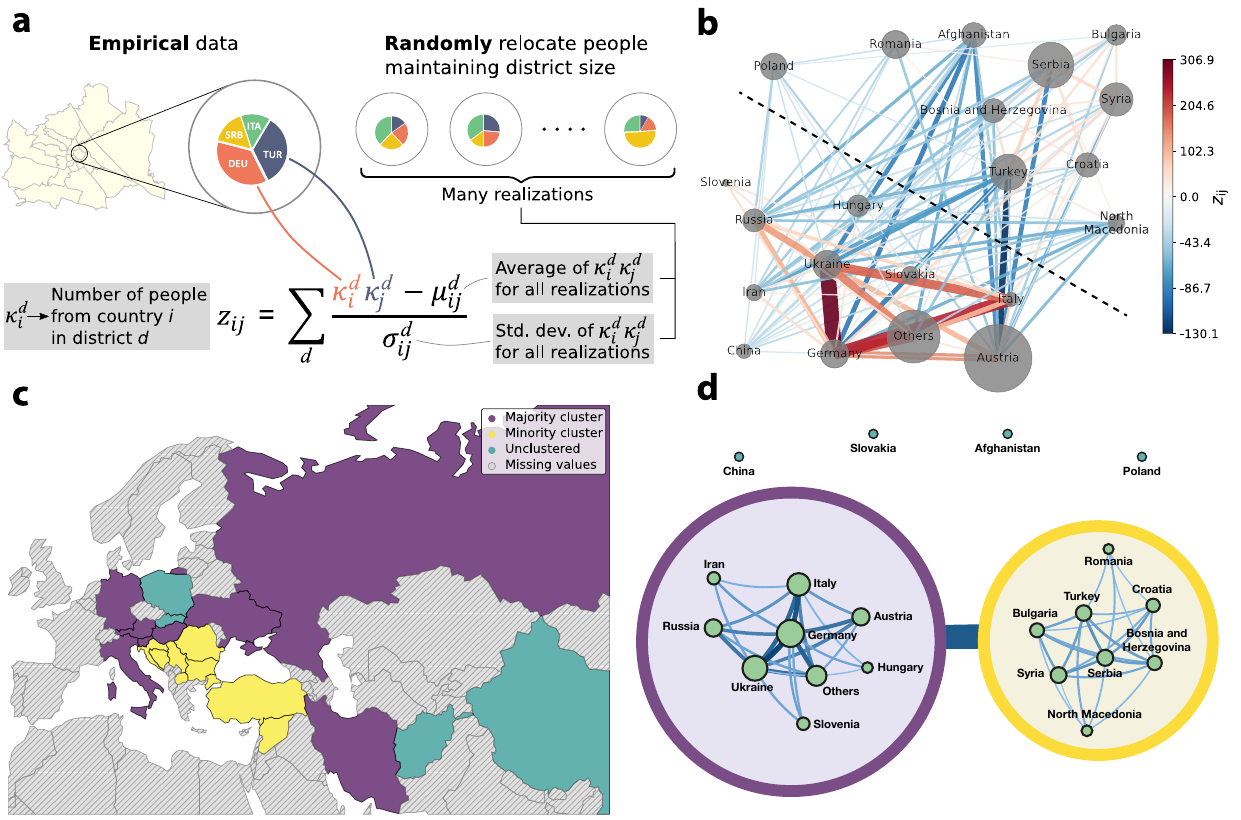}
\caption{\textbf{Clustering nationalities based on district co-residence patterns.} \textbf{a} Diagram of the method used to build the network of co-residence. \textbf{b} $\textit{z}$-score network. Edges with $\lvert z_{ij} \rvert \le 20$ are not shown for better readability. The width and colour transparency of the links are proportional to the $\textit{z}$-score values, while the size of the nodes reflects the total population of each country, except for Austria. The network has been produced using Noverlap layout of Gephi software \cite{gephi}. Two prominent clusters are marked with a dashed line. \textbf{c} World map of district co-residence patterns. Countries that belong to the same cluster have the same colour. The larger cluster, comprising the majority of Vienna's population, is shown in purple and referred to as the ``majority cluster''. The smaller cluster, with fewer residents, is depicted in yellow and termed the ``minority cluster''. Unclustered countries are shown in cyan. \textbf{d} World network of district co-residence patterns. The size of the nodes represents the total $z$-score for each cluster and country. The links represent the connections between nodes obtained from the cluster analysis with Infomap \cite{mapequation2023software}; the thicker the line, the stronger the connection. }
\label{fig:fig2}
\end{figure*}
}

\begin{figure*}[bt!]
\centering
\includegraphics[width=\textwidth]{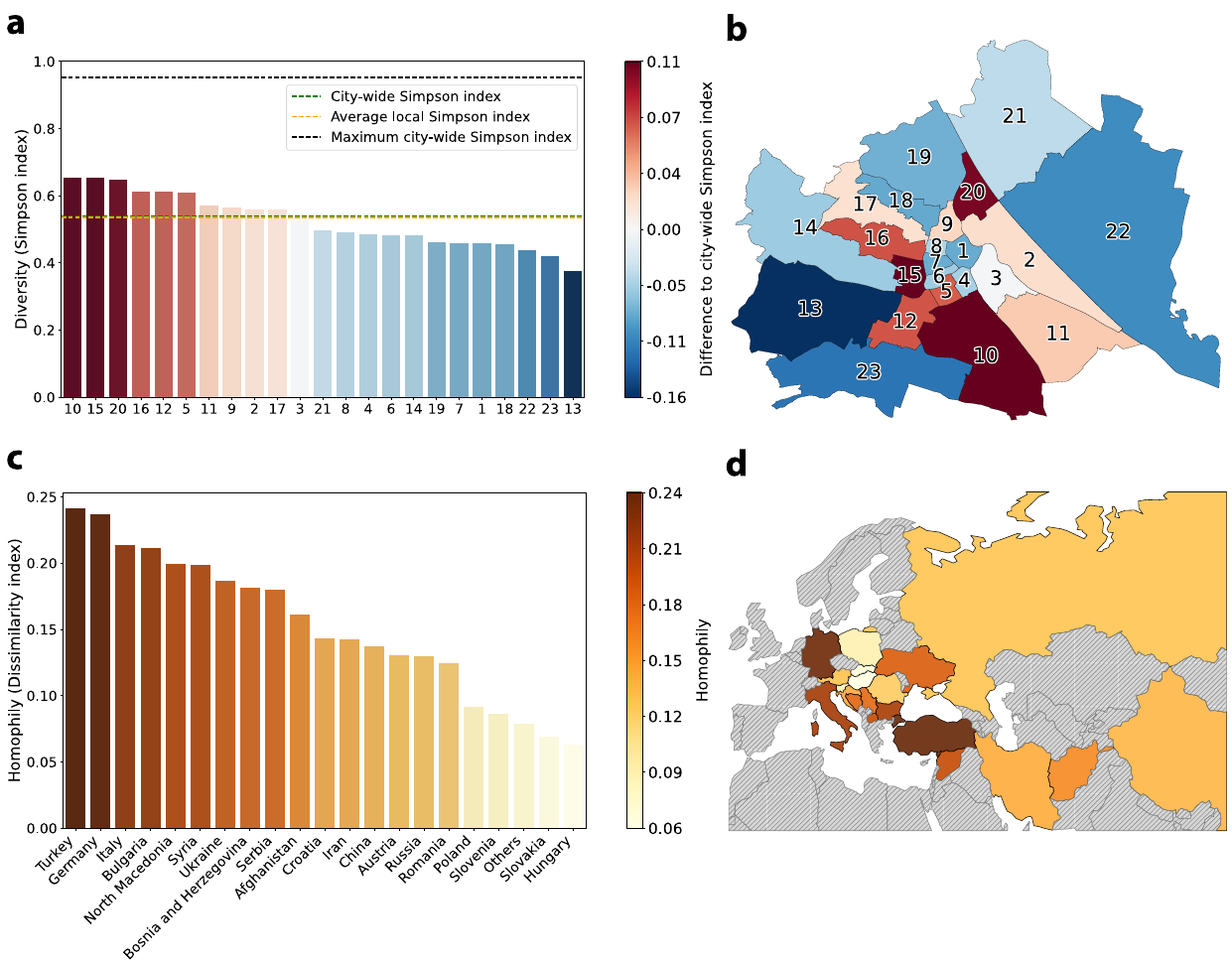}
\caption{\textbf{District diversity and national homophily in Vienna.} \textbf{a} District diversity. Districts are ranked from highest to lowest by estimated diversity in Vienna. \textbf{b} Vienna diversity map. \textbf{c} National homophily. Nationalities are ranked from highest to lowest according to the estimated homophily among countries. \textbf{d} World homophily map.}
\label{fig:fig4}
\end{figure*}

{
\paragraph{Influence of district diversity and national homophily on residential clustering.} We analyse the foreigner diversity within the city's districts, and the degree of country-based homophily among residents. We apply two distinct statistical indices: the Simpson index and the Dissimilarity index. The Simpson index is a measure of concentration that quantifies heterogeneity of the resident population by nationality. A higher Simpson index reflects a more diverse population within a district, indicating a greater variety of nationalities living in close proximity (Fig. \ref{fig:fig4}\textbf{a} and Fig. \ref{fig:fig4}\textbf{b}). On the other hand, the Dissimilarity index is employed to measure homophily, or the tendency of individuals from the same nationality to cluster together within certain districts (Fig. \ref{fig:fig4}\textbf{c} and Fig. \ref{fig:fig4}\textbf{d}). This index helps us assess how much the population of each nationality tends to concentrate in specific areas rather than being evenly distributed across the city \cite{simpson2007ghettos,harris2017measuring}.
}

{
The Simpson index expresses the probability that two randomly selected individuals from the district/city have different nationalities. Calculated for one district \textit{d} of the city, the Simpson index, $S^d$, becomes \cite{duncan1955methodological,white1986segregation}:
\begin{equation}
    S^d=1-\sum_i\left(P_i^d\right)^2,
    \label{eq:Simpson}
\end{equation}
where $P_i^d=N_i^d/N^d$ is the fraction of inhabitants of nationality \textit{i} residing in district \textit{d}. It is a measure between 0 and a maximum value, which depends on the number of distinct nationalities, $C^d$, residing in the district. The higher $S^d$, the more nationality-diverse the district is. A district is fully diverse ($S^d_{\text{max}}=1-1/C^d$) when the population in the district $d$ is equally distributed among all existing nationalities. In contrast, a district has no diversity ($S^d_{\text{min}}=0$) when it is inhabited exclusively by a single nationality. The Simpson index is closely related to entropy, a quantitative measure of disorder (Supplementary Note \ref{sec:entropy}). This relationship is often utilised within segregation studies to assess the unequal distribution of national groups across neighbourhoods or between global and local scales \cite{finney2009population}. 

The city-wide Simpson index can be derived analogously to Eq. (\ref{eq:Simpson}) by using the full population distribution in Vienna. The value for the city of Vienna is $S_\text{City} = 0.541$. This number serves as a reference point for comparing with the average local Simpson index, which is the population-weighted average of the local Simpson index values across all districts \cite{zuccotti2023exploring}. It is computed as:
\begin{equation}
S=\sum_d\left(\frac{N^d}{N_{\text {City}}} S^d\right).
\label{eq:Averagesimpson}
\end{equation}

When the average local Simpson index is substantially lower than the city-wide Simpson index, it indicates a high degree of segregation, with districts tending to be highly homogeneous, often dominated by a single nationality. In Vienna, the average local Simpson index is $S = 0.535$. This implies that there is approximately a 54\% chance that a randomly selected pair of individuals from the same district belong to different nationalities. This value closely aligns with the probability observed when sampling a pair randomly from the entire city, suggesting that overall segregation levels in Vienna are relatively low (Fig. \ref{fig:fig4}\textbf{a}). The Simpson index is highest for district 10 and lowest for district 13 (Supplementary Table \ref{tab:tabs3} for detailed values). The percentage of foreigners is correlated with the diversity (the Supplementary Note \ref{sec:diversity}). It might initially seem that this pattern arises inherently from the definition of the Simpson index. However, in a hypothetical scenario in which all foreigners in a specific district belong to a single nationality, the percentage of foreigners would be high, yet diversity would remain low. Therefore, this finding confirms that foreigners are distributed relatively homogeneously across districts, residing in the more diverse ones and with no evidence of a specific foreigner group forming a concentrated enclave in any particular district (Fig. \ref{fig:fig1}\textbf{b} and Supplementary Fig. \ref{fig:figs4}).
}

{
To define national homophily, the tendency to reside among others of the same nationality \textit{i}, we calculate the Dissimilarity index $D_i$, which indicates the percentage of members of that group who would need to relocate so that the distributions at the district-level match the distribution at the city-level \cite{zuccotti2023exploring}:
\begin{equation}
D_i=\frac{\sum_d \frac{N^d}{N_{\text {City}}} \cdot\left|P_i^d-P_i\right|}{2 P_i\left(1-P_i\right)}.
\label{eq:dissimilarity}
\end{equation}

This index ranges from 0 to 1. A nationality is fully segregated ($D_i=1$) when all inhabitants of that origin reside in the same district. Conversely, a nationality is unsegregated ($D_i=0$) when its population is evenly distributed across all districts (Fig. \ref{fig:fig4}). The Dissimilarity index is highest for Turkish nationals and lowest for Hungarians (Supplementary Table \ref{tab:tabs4}). These results reveal notable differences in how national groups are distributed across Vienna's districts. Such differences may reflect a combination of factors, including budgetary constraints, shared language, traditions, social norms, and existing community structures. Over time, these group-level residential concentrations can contribute to broader patterns of district-level segregation in the city. For instance, even within clusters associated with geographic proximity, the concentration of a single nationality in particular districts may reinforce segregation patterns and contribute to the formation of distinct spatial and cultural residential environments. Understanding this interplay is essential, as it highlights the multi-layered nature of clustering, in which nationality-specific residential distributions and broader group-level patterns jointly shape the observed residential structure. We have also examined other indices using different statistical techniques in Supplementary Note \ref{sec:othermeasures}.
}

\section*{Results}\label{sec:results}
\subsection*{Determinants of co-residence clusters}

{
\paragraph{Geographical and national homophily.}
The key question is what factors determine the segregation of countries into two clusters. The world map of district co-residence patterns (Fig. \ref{fig:fig2}\textbf{c}) suggest that geographic proximity could be a key factor in the division of countries \cite{crang2013cultural}. Furthermore, the world network of district co-residence patterns (Fig. \ref{fig:fig2}\textbf{d}) reveals that connections are typically stronger between geographically proximate countries, further supporting this hypothesis. Nationalities grouped within the same cluster often originate from nearby regions and may share aspects of cultural background, sharing similar languages, traditions, and social norms. This geographical homophily may drive the preference for foreigners from these countries to live in close proximity, reinforcing the observed segregation.
}

{
The five most populous nationalities in the larger cluster are Austria, Ukraine, Germany, Russia, and Hungary, while the most populous nationalities in the smaller cluster are Serbia, Turkey, Syria, Romania, and Croatia. Most countries within each cluster belong to geographically proximate or regionally connected areas. This does not imply that every pair of countries within a cluster is directly adjacent or geographically close. Rather, the clusters can reflect indirect regional continuity, where countries are connected through intermediate countries within the same group that bridge nearby regions. This pattern suggests that geographical homophily is associated with the clustering structure, although other factors, including economic conditions, may also contribute to the observed district-level residence patterns. Among the unclustered countries, China and Afghanistan stand out for their relatively distant geographic locations compared to the others.
}

{
Beyond geographical homophily at the macro level, national homophily, understood here as the tendency of residents from the same nationality to be concentrated in certain districts, may also contribute to the observed residential sorting patterns within Vienna. In the Methods section, we characterise the homophily level for each nationality using the Dissimilarity index, defined as the proportion of members of a group who would need to relocate for the group fractions at the district-level to match the city-wide distribution. Turkey, Germany, and Italy have the highest levels of homophily, while Hungary and Slovakia have the lowest.  
}

{
When a particular nationality exhibits strong homophily, with a large share of its residents concentrated in specific districts, it may act as a residential anchor within the shared residence network, being associated with stronger shared residence links to other nationalities with related cultural or regional backgrounds \cite{PrietoCuriel2024}. Such concentrations can contribute to broader patterns of cultural and regional clustering across the city. Therefore, national homophily at the district level may reinforce the clustering structure observed among larger groups of countries. Next, we examine how neighbourhood income disparities and district diversity are associated with these residential patterns.
}

{
\paragraph{Neighbourhood income.} Neighbourhood income is closely associated with residential patterns and the socio-economic composition of urban areas. By examining income disparities across districts, we assess whether the residence clusters identified in the network are systematically related to the economic profile of the districts in which different nationalities are registered. To do this, we incorporate an additional publicly available dataset to approximate district-level income (Supplementary Note \ref{sec:incomerepresentativeness}).
}

{
We approximate the income distribution at the district level by the average net yearly income of the inhabitants of the district (Supplementary Note \ref{sec:incomerepresentativeness}). Then, for each country, we compute the Pearson correlation coefficient between the proportion of the country's population in a district and the district's average net income. By expressing nationality-specific counts as population fractions at the district level, our analysis partially accounts for differences in housing availability and residential capacity across districts (Supplementary Notes). A high positive correlation indicates that the country's residents are concentrated in wealthier districts, suggesting greater affluence. Conversely, a strong negative correlation implies that its residents tend to live in lower-income districts. This analysis quantifies the tendency of individuals from a specific country to reside in districts in which the perceived socio-economic status of other residents, or their own, is associated with higher or lower average income. 
We then rank the countries from highest to lowest Income-Population fraction correlation (Fig. \ref{fig:fig3}\textbf{a}). Austria exhibits a strong positive correlation, indicating that Austrians are overrepresented in higher-income districts (positive association between district income and population fraction). In contrast, Bosnia and Herzegovina shows a negative correlation, indicating that its residents are overrepresented in lower-income districts.
}

{
There is an economic division in Vienna between the majority and the minority clusters (Fig. \ref{fig:fig3}\textbf{a}). Nationalities in the majority cluster tend to be overrepresented in districts with higher average incomes, whereas nationalities in the minority cluster tend to be overrepresented in lower-income districts. Hungary and \textit{Others} are exceptions within the majority cluster, as they also exhibit negative income-population fraction correlations. These results indicate that district-level income disparities are closely aligned with the shared residence clusters identified in the network. Supplementary Note \ref{sec:rental} explores an alternative perspective on neighbourhood economic disparities, presenting consistent outcomes and complementary analyses based on a dataset of district-level housing rental prices in Vienna.

\begin{figure*}[bt!]
\centering
\includegraphics[width=\textwidth]{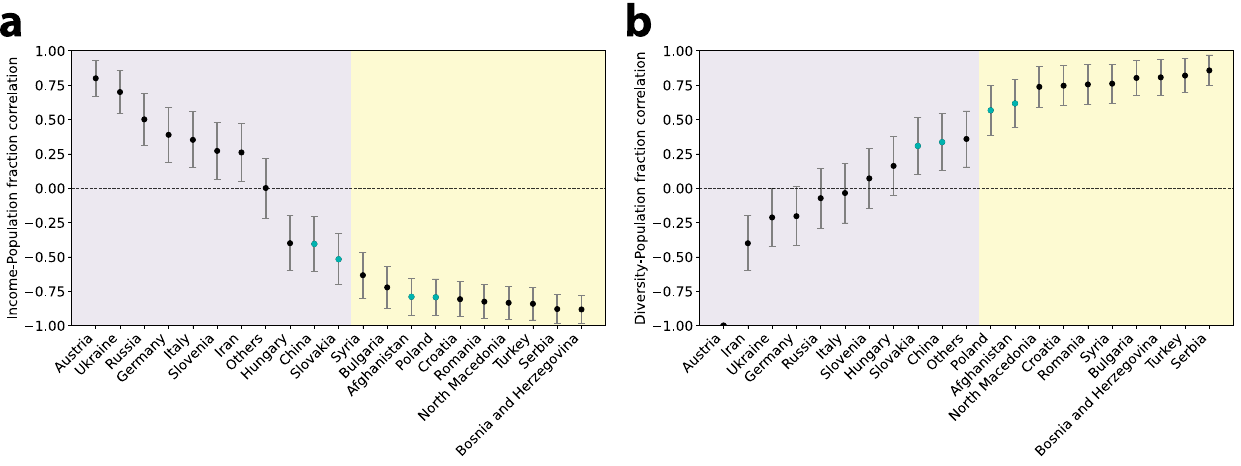}
\caption{\textbf{Socio-economic and diversity factors underlying nationality clusters.} \textbf{a} Correlation between district income and population fraction for each nationality. \textbf{b} Correlation between diversity and population fraction for each nationality. Nationalities are ranked by their correlation values, from highest to lowest for income-population fraction correlation in \textbf{a}, and from lowest to highest for diversity-population fraction correlation in \textbf{b}. The nationality clusters are colour-coded consistently with Fig. \ref{fig:fig2}\textbf{c}. 
}
\label{fig:fig3}
\end{figure*}
}

{
\paragraph{Neighbourhood diversity.} To estimate urban diversity, we introduce the Simpson index, which can be calculated for each district (local) and the whole city. It expresses the probability that two randomly selected individuals from the district (or the city) have different nationalities. After calculating district-level diversity, we compute the Pearson correlation coefficient for each country, examining the relationship between a country's population fraction in a district and the district's diversity. In this context, the analysis measures the extent to which the population of a specific country experiences diversity within the urban environment. The two identified clusters are distinctly separated with respect to diversity (Fig. \ref{fig:fig3}\textbf{b}). Populations from countries within the majority cluster tend to reside in districts characterised by lower diversity, whereas those in the minority cluster tend to be registered in more diverse districts.. Countries with correlations near zero indicate that their populations are equally likely to reside in districts with either higher or lower levels of diversity. 
}

{
Diversity within a district may reflect broader socio-economic and cultural factors, such as income, educational background, foreign-born populations, and geographic proximity. Wealthier districts may attract a specific mix of nationalities due to economic constraints that limit where foreigners can afford to live, thereby decreasing diversity in these areas, while less affluent districts may exhibit higher diversity, resulting in a more heterogeneous population. In this context, the grouping of nationalities into two distinct groups may be due to income inequalities and differences in geographical proximity, as reflected in patterns of residential segregation and district-level diversity. These associations may also contribute to self-reinforcing residential patterns: as certain nationalities concentrate in particular districts, their presence can shape the demographic composition of those areas, either attracting more foreigners from similar backgrounds and reinforcing homogeneity or encouraging further diversity and integration.
}


\section*{Discussion} \label{sec:discussion}

{
Our analysis reveals a pattern of district-level residential sorting within the Vienna population, characterised by two distinct groups of nationalities. Citizens from the same cluster tend to be registered in the same districts or adjacent ones, whereas those from different clusters are less likely to share the same district-level residential environments. These clustering patterns are associated with neighbourhood income and diversity. In particular, citizens from nationalities in the majority cluster are more prevalent in districts with higher average incomes, whereas other groups are more concentrated in districts with lower average incomes. At the same time, district diversity is closely aligned with these patterns, with the two clusters differing systematically in the diversity of the districts where their residents are registered. These results suggest that economic conditions and district composition are important dimensions of nationality-based residential sorting in Vienna.

The observed patterns likely arise from interactions among several mechanisms, some of which cannot be disentangled using a static, district-level snapshot of residence registrations, such as whether housing relocation reinforces existing residential patterns over time. Yet, by normalizing nationality-specific counts to the district population, our approach implicitly controls for differences in district size and housing capacity, thereby partially accounting for housing availability across districts. Moreover, our supplementary analysis using district-level rental prices highlights the role of housing-market conditions as an additional dimension associated with residential outcomes. Together, these results suggest that segregation patterns in Vienna reflect an interplay between multiple factors and constraints, rather than the effect of a single dominant mechanism.
}

{
The relationship between district diversity and the clustering of country populations in Vienna is a complex, interdependent process in which diversity can act both as a cause and a consequence of segregation, generating a multifaceted feedback loop. On the one hand, diversity can influence the residential tendencies of different nationalities, with some residents attracted to more or less diverse areas. This mechanism of self-segregation, or the tendency of individuals to self-select into places based on nationality or geographical proximity, drives the formation of more homogeneous clusters, reinforcing patterns of low diversity in certain districts. 
}

{
Overall, migration and residential segregation are path-dependent social processes shaped by societal and economic forces, including housing prices, social housing allocation, and labour-market geography. The observed patterns of residential segregation are therefore not solely the product of geographical proximity or economic factors, but likely reflect the interplay between individual-level constraints, community structures, housing-market conditions, and broader systemic influences. This highlights segregation as a multifactorial phenomenon in which district-level residential clustering emerges from overlapping social, economic, and institutional dimensions.
}

{
The dynamics of segregation are a key component for understanding the shape and scale of internal migration, particularly in relation to urbanisation and mobility flows between human settlements \cite{zuccotti2023exploring}. Yet, disentangling residential segregation arising from housing relocation from that arising from the arrival and departure of people in the city remains an ongoing challenge. Future work could extend this district-level analysis by incorporating finer spatial units, longitudinal residence histories, mobility data, municipal housing, urban amenities, vulnerability indicators, cultural distance metrics, and infrastructure accessibility. Such extensions would help clarify how district-level residential sorting relates to finer-scale segregation, lived urban experience, and changes in residential patterns over time.
}

\section*{Data availability} 

The raw and processed data are not available due to privacy laws. The Federal Ministry of the Interior of Austria safeguarded the dataset and made it accessible to our research institution in accordance with strict data protection regulations. Researchers must reach individual agreements with the Federal Ministry of the Interior of Austria to access this data.

In a repository, we provide the aggregate statistics on the proportion of citizens of each nationality living in each district \cite{github}. Together with the publicly available complementary datasets, these data are sufficient to reproduce all the analyses performed in this study.

\section*{Code availability}
The code developed to perform all analyses is available in a GitHub repository \cite{github}.


\section*{Acknowledgments}
M.S. was partly funded by MCIN/AEI/ 10.13039/501100011033 and by ``ESF Investing in your future'', grant number PRE2020-093266; by MCIN/AEI/ 10.13039/501100011033, by ``ERDF A way of making Europe'', grant number PID2022-140757NB-I00, and the support of Generalitat de Catalunya, grant number 2021SGR00856. O.A. and R.P.C. were funded by the Austrian Federal Ministry for Climate Action, Environment, Energy, Mobility, Innovation and Technology (2021-0.664.668) and the Austrian Federal Ministry of the Interior (2022-0.392.231). F.K., S.M.G., and A.M.J. were partly funded by the EU Horizon Europe project MAMMOth (Grant Agreement 101070285). S.M.G. was supported by the Austrian research agency (FFG) under project No. 873927 ESSENCSE, and by grant PID2024-157869NB-I00, funded by MICIU/AEI/10.13039/50110001103.

\section*{Author contributions statement}

M.S. performed the analysis and developed models and simulations. M.S., S.M.G., A.M.J., and F.K. defined the problem and designed the solution and algorithms. O.A. and R.P.C. collected and processed the data. All authors designed research, analysed results, discussed results, and wrote and reviewed the manuscript.

\section*{Competing interests}
The authors declare no competing interests. 

\section*{Ethical approval}
This article does not contain any studies with human participants performed by any of the authors.

\section*{Informed consent}
This article does not contain any studies with human participants performed by any of the authors.

\end{document}


\maketitle

\tableofcontents
\newpage
\listoffigures
\listoftables
\newpage

\section{Residence data} \label{sec:residencedata}

\subsection{Population representativeness}

Over the past six decades, Vienna has undergone some demographic shifts. What was once a stagnating urban center became a shrinking city and eventually a rapidly expanding metropolis. Simultaneously, an aging population has given way to a younger and more vibrant demographic landscape, largely driven by international migration \cite{suitner2021vienna,kral_building_2021}. Several events have contributed to Vienna's significant population growth, including the fall of the Iron Curtain, the wars in the former Yugoslavia, Austria's EU accession, and the EU enlargements in 2004, 2007, and 2013. Additionally, migration from conflict regions like Syria, Afghanistan, and Ukraine has played a key role. Since joining the EU in 1995, Vienna's population has increased by more than 439,000 people, driven by migration and a positive net balance of births and deaths. By early 2023, Vienna's population reached 1.9 million \cite{fassmann2008austria,bischof2017migration}. The impact of young foreigners is evident in Vienna's population composition. As of early 2023, according to the official website of the city of Vienna, 34.2\% of residents held foreign citizenship, 39.3\% were born abroad, and 44.4\% had foreign origins, meaning they either held foreign citizenship or were Austrian citizens born abroad \cite{migrantsvienna}.

Our primary dataset is sourced from the Austrian government and provides detailed residential information for all foreign citizens residing in Austria, aligning with Vienna's definition of foreign citizenship \cite{migrantsvienna}. This information is derived from the mandatory residence registration form (“Meldezettel”) and has been systematically recorded since November 2022. For this study, we utilized a snapshot capturing the residence locations of the entire foreign population across Vienna's districts on 22nd September 2023 (see Supplementary Fig. \ref{fig:figs1}\textbf{a}). To address the absence of Austrian population data in our primary dataset, we used official district-level population counts from an external source \cite{viennapopulation} (Supplementary Fig. \ref{fig:figs1}\textbf{b}). To estimate the Austrian population in each district (Supplementary Fig. \ref{fig:figs1}\textbf{c}), we calculated the difference between the total population figures (Supplementary Fig. \ref{fig:figs1}\textbf{b}) and the number of registered foreigners recorded in our dataset (Supplementary Fig. \ref{fig:figs1}\textbf{a}). This approach maintains alignment with official statistics while ensuring accurate representation of district-level population distributions. The ratio of foreign citizens to local residents, illustrating the proportion of foreigners in each district relative to Austrians (Supplementary Fig. \ref{fig:figs1}\textbf{d}). This measure provides insights into the extent to which local residents are exposed to or interact with non-local individuals, varying across Vienna's districts (Fig. \ref{fig:fig1}\textbf{a}).

Our dataset serves as a robust foundation for analysing district-level shared residence patterns among foreigners and nationals in Vienna. By uncovering two distinct clusters shaped by income disparities, district diversity, and nationality-driven homophily, this study offers valuable insights into the complex dynamics of urban segregation and integration. These findings illuminate the interplay between socio-economic factors and geographical proximity tendencies, revealing how they collectively influence residential distribution in a multicultural urban environment. Through this lens, we gain a deeper understanding of the mechanisms that drive both separation and cohesion within Vienna's diverse communities.
\begin{figure}[bt!]
\centering
\includegraphics[width=\textwidth]{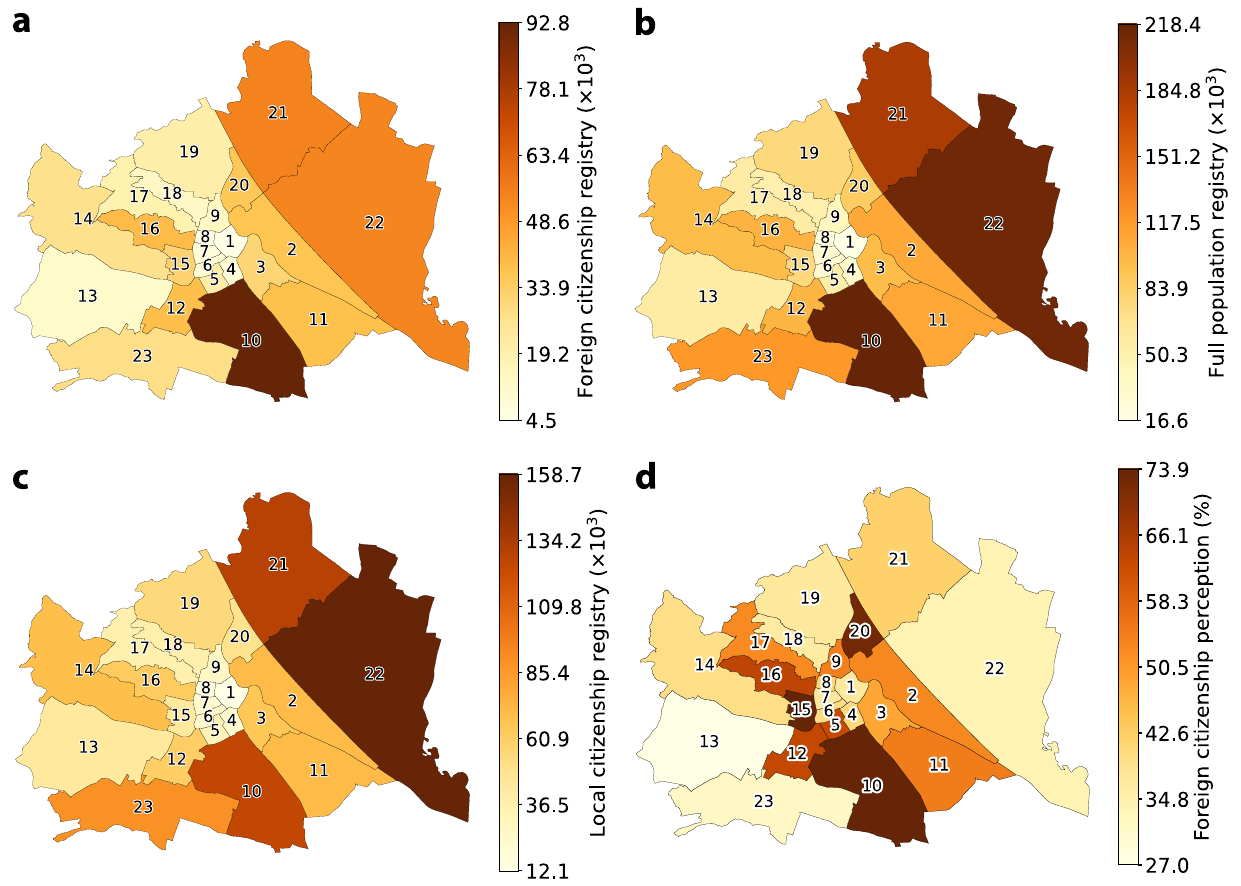}
\caption[Demographics of Vienna across citizenship categories.]{\textbf{Demographics of Vienna across citizenship categories.} \textbf{a} Distribution of residents with foreign citizenship based on official registry data. \textbf{b} Total population distribution across Vienna's districts. \textbf{c} Distribution of residents with Austrian citizenship inferred from supplementary data. \textbf{d} Proportion of foreign citizens relative to local residents, capturing the perceived presence of foreign populations in each district. Data is presented in Supplementary Table \ref{tab:tabs1}, where districts are sorted from highest to lowest foreign citizenship perception.}
\label{fig:figs1}
\end{figure}

\begin{table}[bt!]
\centering
\caption[District-Level citizenship data]{\textbf{District-Level citizenship data}. Numerical data on foreign and local citizenship populations by district on 22nd September 2023, corresponding to the map visualisations represented in Supplementary Fig. \ref{fig:figs1}. Districts are ranked from highest to lowest by the percentage of foreign citizenship.}
\begin{tabular}{c|c|c|c|c|c}
\begin{tabular}{c}
\textbf{District} 
\end{tabular}  & \begin{tabular}{c}
\textbf{Full} \\ \textbf{population} \\ \textbf{registry}
\end{tabular} & \begin{tabular}{c}
\textbf{Foreign} \\ \textbf{citizenship} \\ \textbf{registry} 
\end{tabular} & \begin{tabular}{c}
\textbf{Local} \\ \textbf{citizenship} \\ \textbf{registry}
\end{tabular} & \begin{tabular}{c}
\textbf{Foreign} \\ \textbf{citizenship} \\ \textbf{perception (\%)}
\end{tabular} & \begin{tabular}{c}
\textbf{Foreign} \\ \textbf{citizenship} \\ \textbf{percentage (\%)}
\end{tabular}\\
\hline
\hline
15 & 76\,109 & 32\,349 & 43\,760 & 73.92 & 42.50 \\ 
10 & 218\,415 & 92\,767 & 125\,648 & 73.83 & 42.47 \\ 
20 & 85\,690 & 35\,997 & 49\,693 & 72.44 & 42.01 \\ 
16 & 102\,444 & 39\,854 & 62\,590 & 63.67 & 38.90 \\ 
12 & 100\,281 & 38\,806 & 61\,475 & 63.12 & 38.70 \\ 
5 & 55\,018 & 21\,203 & 33\,815 & 62.70 & 38.54 \\ 
11 & 109\,038 & 38\,483 & 70\,555 & 54.54 & 35.29 \\ 
9 & 42\,206 & 14\,845 & 27\,361 & 54.26 & 35.17 \\ 
2 & 108\,269 & 37\,445 & 70\,824 & 52.87 & 34.59 \\ 
17 & 56\,033 & 19\,303 & 36\,730 & 52.55 & 34.45 \\ 
3 & 96\,756 & 31\,690 & 65\,066 & 48.70 & 32.75 \\
21 & 183\,895 & 54\,576 & 129\,319 & 42.20 & 29.68 \\
8 & 24\,674 & 7\,245 & 17\,429 & 41.57 & 29.36 \\
4 & 33\,633 & 9\,762 & 23\,871 & 40.89 & 29.03 \\
6 & 31\,423 & 9\,014 & 22\,409 & 40.22 & 28.69 \\
14 & 96\,828 & 27\,570 & 69\,258 & 39.81 & 28.47 \\
1 & 16\,620 & 4\,524 & 12\,096 & 37.40 & 27.22 \\
19 & 75\,517 & 20\,533 & 54\,984 & 37.34 & 27.19 \\
7 & 31\,581 & 8\,553 & 23\,028 & 37.14 & 27.08 \\
18 & 51\,559 & 13\,784 & 37\,775 & 36.49 & 26.73 \\
22 & 212\,658 & 54\,007 & 158\,651 & 34.04 & 25.40 \\
23 & 117\,882 & 28\,539 & 89\,343 & 31.94 & 24.21 \\
13 & 55\,568 & 11\,817 & 43\,751 & 27.01 & 21.27 \\
\end{tabular}
\label{tab:tabs1}
\end{table}

\subsection{Supplementary datasets} \label{sec:otherdatasets}
In addition to the other datasets described in the Methods section, we used several supplementary datasets to extend our analysis and provide further context for the findings in this Supplementary Information. Net income data by country in diverse year ranges were sourced from the World Population Review website \cite{incomecountries}, which integrates information from the World Bank Group, Eurostat, Giving What We Can, and Gallup \cite{incomecountries1, incomecountries2, incomecountries3, incomecountries4}. The specific column used provides median income values in international dollars (\textdollar). To facilitate standardised comparisons, we employed the average exchange rate for 2020, obtained from the source Exchange Rates UK \cite{exchangerates2020}, using a conversion factor of \textdollar1 USD = \texteuro0.877. Additionally, data on average rental prices per square meter for Vienna districts, sourced from a detailed analysis of housing costs \cite{viennarental}, was incorporated to investigate economic disparities and their influence on residential clustering patterns. The column used here was the average per square meter. These supplementary datasets enrich our primary data, enabling a deeper examination of the socio-economic and demographic factors shaping residence segregation in Vienna.

\subsection{Income and Rental representativeness} \label{sec:incomerepresentativeness}
We approximate that the entire population residing within a district earns the district’s average net income. This assumption is a standard approach in many scientific studies aiming to link economic indicators to spatial or demographic patterns \cite{moro2021mobility}. Using the district-wide average yields a consistent, interpretable metric that reflects the general economic conditions across districts. This approach is particularly effective for large-scale analyses in which individual-level income data are unavailable or impractical to obtain, as it captures the aggregate economic environment of a district. Furthermore, district-level average net income data, filtered for the year 2020, serve as a proxy for economic status \cite{viennaincome}. Although these data do not align perfectly with the time frame of the residence registration data (2023), they are the most recent available dataset. Using 2020 as a reference year is a reasonable approximation, given the typically gradual changes in average income over short periods, unless major economic disruptions are present. This approach allows us to estimate relative income variations across districts and their potential impact on residential segregation patterns. 

\begin{figure}[bt!]
\centering
\includegraphics[width=\textwidth]{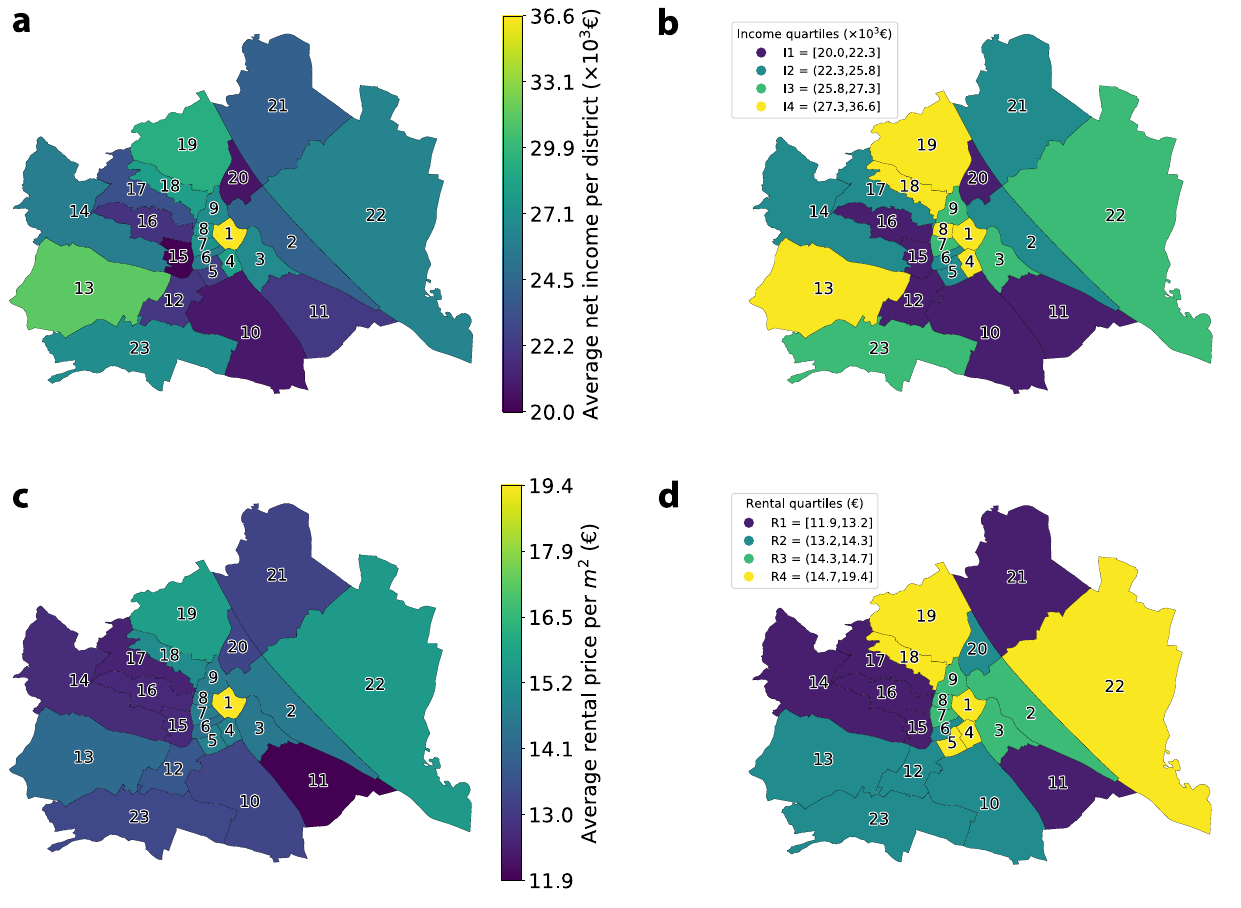}
\caption[Average net Income and Rental prices in Vienna districts.]{\textbf{Average net Income and Rental prices in Vienna districts.} \textbf{a} Spatial distribution of average net income across districts. Colour bar is in log scale. \textbf{b} Districts categorised into quartiles based on their average net income. \textbf{c} Spatial distribution of average rental prices per square meter across districts. The colour bar is in log scale. \textbf{d} Districts categorised into quartiles based on their average rental prices.}
\label{fig:figs2}
\end{figure}

For rental prices, we adopt a similar district-level framework by using the average rental price per square meter in each district \cite{viennarental}. This dataset serves as an economic indicator that complements the income data, providing insights into housing costs and their interplay with residential clustering. While individual variations in rental agreements exist, the average price per square meter provides a general measure of housing affordability and economic accessibility across different districts. This district-level aggregation aligns with our income analysis, providing consistency in scale and methodology. Although these district-level approximations introduce some limitations, they remain an effective tool for analysing urban socio-economic patterns. They provide a high-level understanding of the relationships between income, housing, and residential tendencies in Vienna, setting the stage for further research to refine these insights with more granular data (Supplementary Fig. \ref{fig:figs2}). The subpanels \textbf{a} and \textbf{c} depict the average net income and the average rental prices as district-level maps, providing a clear spatial representation of economic disparities within the city. To facilitate comparative studies and quartile-based analyses, we further compute and map the income and rental price quartiles for Vienna’s districts in subpanels \textbf{b} and \textbf{d}. These quartile classifications provide a deeper understanding of the socio-economic factors that influence residential patterns across the city.

\section{Socio-Demographic insights of Vienna} 
Vienna is divided into 23 diverse districts, each contributing to the city's rich demographic mosaic. As of late September 2023, the city's population consisted of 650 thousand residents with foreign citizenship and 1.3 million with Austrian citizenship, resulting in an average foreign population share of 32.9\%. To ensure data privacy, residential data were aggregated, and only nationalities with at least 10 individuals per district were included. 

The primary countries of origin for residents with foreign backgrounds have remained largely consistent in recent years. In September 2023, 91 thousand individuals held Serbian citizenship, 57 thousand had Turkish citizenship, and 48 thousand were Syrian nationals. The districts with the highest proportions of residents with foreign citizenship were Rudolfsheim-Fünfhaus (15th district), Favoriten (10th district), and Brigittenau (20th district), each with nearly 43\% of their populations being foreigners. Conversely, Hietzing (13th district) had the lowest proportion, with only about 21\% of its residents holding foreign citizenship (Supplementary Table \ref{tab:tabs1}). 

\begin{figure}[bt!]
\centering
\includegraphics[width=\textwidth]{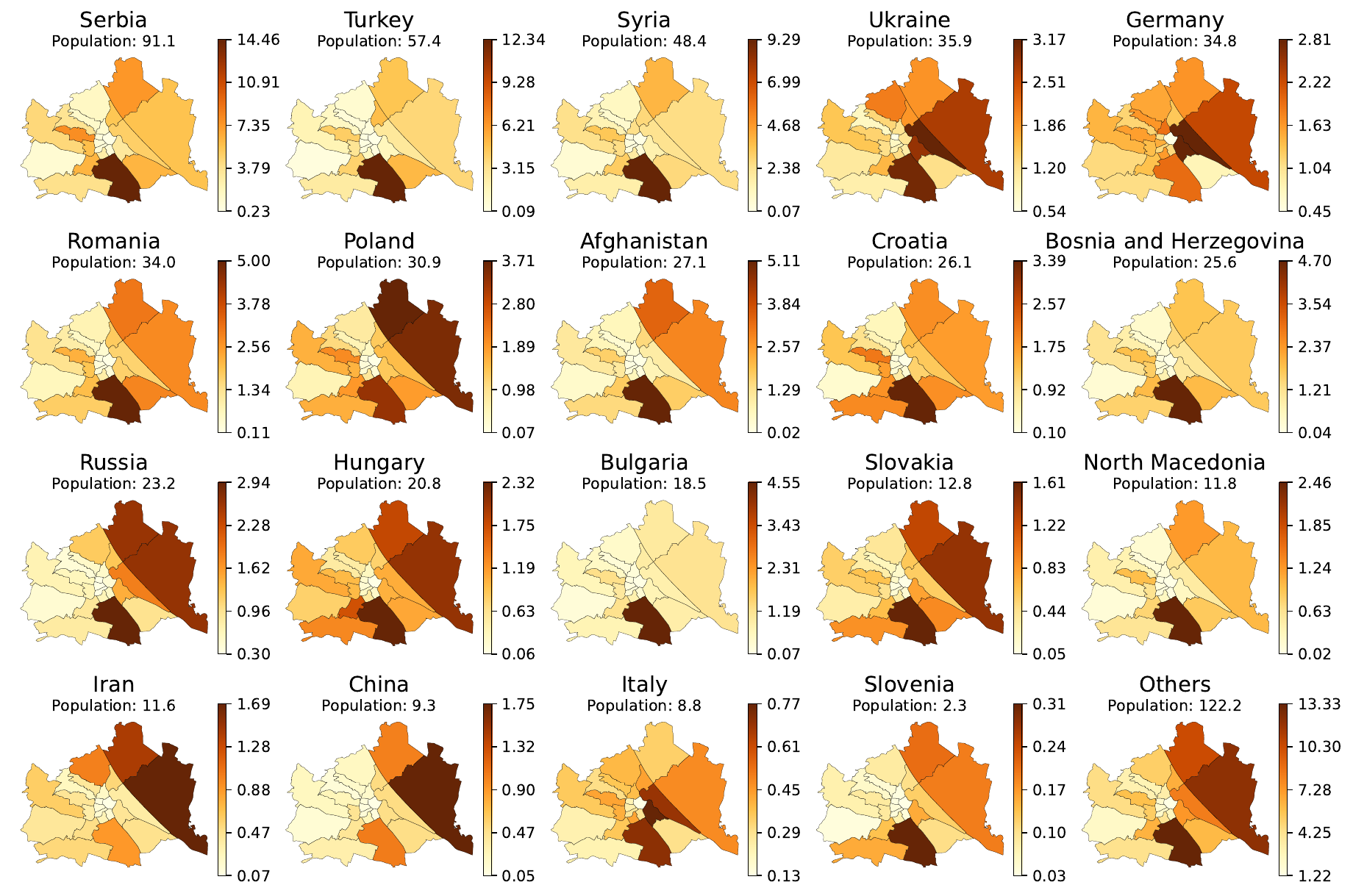}
\caption[Total population distribution of each nationality across Vienna's districts.]{\textbf{Total population distribution of each nationality across Vienna's districts.} Nationalities are ordered in descending order of foreign population in Vienna, with \textit{Others} included at the end. Total population values and colour bar scales are represented in units of $\times 10^3$.}
\label{fig:figs3}
\end{figure}

\begin{figure}[bt!]
\centering
\includegraphics[width=\textwidth]{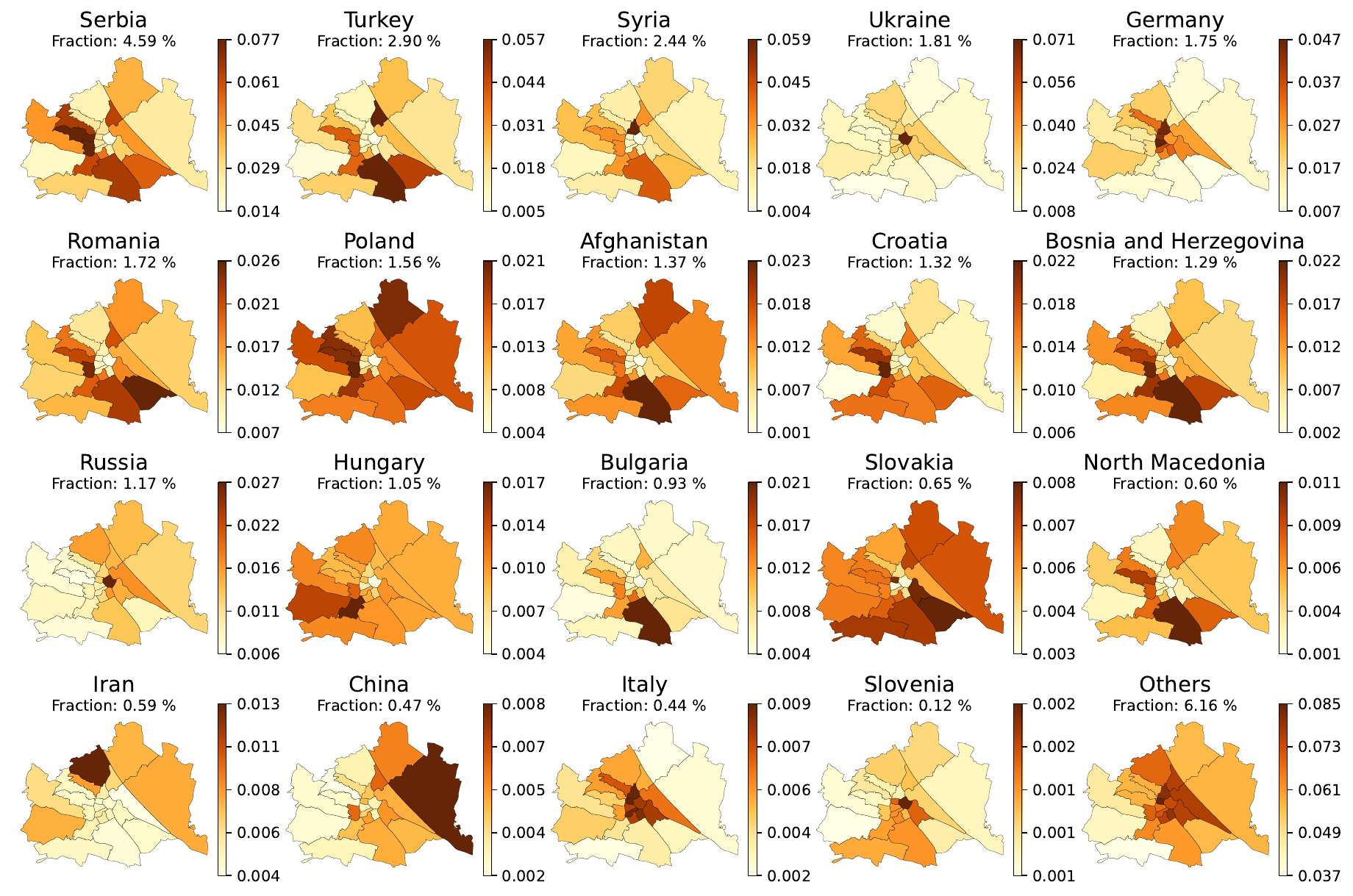}
\caption[Population fraction of each country relative to the whole population in each district.]{\textbf{Population fraction of each country relative to the whole population in each district.} Nationalities are ordered in descending order.}
\label{fig:figs4}
\end{figure}

Districts with greater housing availability attract larger populations from all nationalities (Supplementary Fig. \ref{fig:figs1}\textbf{b}). To address this limitation, we analyse the population fraction of each nationality relative to the total population in each district (Supplementary Fig. \ref{fig:figs4}). This metric normalises the data by accounting for district-level housing availability, making it a more insightful measure for segregation studies. By controlling for district size, it highlights the relative concentration and exposure of each nationality within specific districts. Comparing Supplementary Fig. \ref{fig:figs3} with Supplementary Fig. \ref{fig:figs4}, substantial discrepancies become evident. The normalised population fractions in Supplementary Fig. \ref{fig:figs4} illustrate emerging patterns of segregation and disparities by national origin, providing deeper insights into residential clustering.

\begin{figure}[bt!]
\centering
\includegraphics[width=\textwidth]{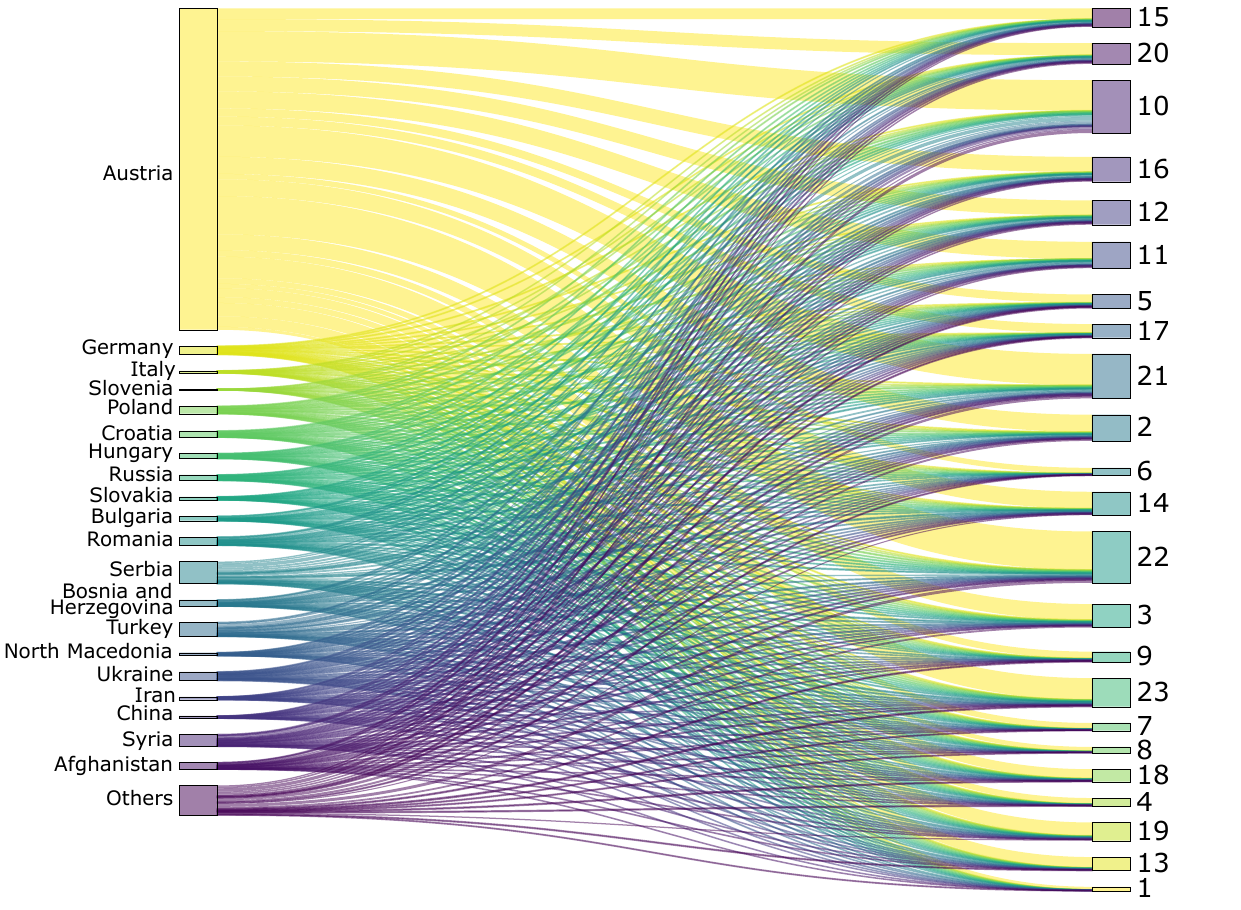}
\caption[Bipartite Network representation.]{\textbf{Bipartite Network representation.} Nationalities are ranked by descending median income in their country of origin \cite{incomecountries}, while districts are ordered by ascending average net income \cite{viennaincome}. The sizes of nodes and links are scaled proportionally to their respective values. The category \textit{Others} is included at the end for completeness.}
\label{fig:figs5}
\end{figure}

Residential patterns can also be conceptualised as a bipartite network, with nationalities forming one set of nodes and districts the other (Supplementary Fig. \ref{fig:figs5}). This representation provides a clear foundation for understanding co-residence tendencies, which are derived by projecting the bipartite network onto the side of nationalities. In this projection, two countries are connected if they share at least one district, with the strength of the connection equal to the number of shared districts. While all nationalities in our dataset share districts with others, the key challenge is identifying which connections are statistically significant beyond chance. This question is addressed in the following section.

\section{Mapping district-level shared residence patterns} \label{sec:coresidence}

Mapping district-level shared residence patterns using the total number of people of different nationalities living across Vienna's districts provides a valuable framework for understanding patterns of residential clustering and integration. This approach enables us to quantify the extent to which individuals of different nationalities share the same district spaces, offering insights into the social dynamics of multicultural urban areas. By analysing these patterns, we can uncover the degree of interaction or separation between communities, which in turn sheds light on broader phenomena such as social cohesion, segregation, and exposure to diversity.

We approximate a country’s interest profile by analysing the collective residential behaviours of its citizens residing in Vienna. However, it is important to note that these patterns are specific to Vienna and may not accurately reflect the broader residential patterns of the country’s population. To obtain a more statistically robust and generalisable understanding, future research would need to compare residential behaviours across multiple multicultural cities worldwide. This broader comparison could yield deeper insights into the dynamics of co-residence tendencies globally.

\subsection{Extracting significant shared residence links}
The methodology outlined in Fig. \ref{fig:fig2}\textbf{a} and the Methods section of the main text provides a framework for identifying significant deviations from random distributions. In this section, we further explore these relationships, presenting robust methodologies for mapping and interpreting district-level shared residence patterns with greater precision and depth.

In Supplementary Fig. \ref{fig:figs6}\textbf{a}, we present the computed \textit{z}-scores matrix, with countries arranged according to the Infomap clustering results (without Bonferroni correction). This matrix provides a more insightful view of the co-residence tendencies than a network representation. It clearly highlights two prominent clusters in the data, reflecting distinct patterns of nationalities that co-occur more frequently than expected by chance. By organising countries into these clusters, we can more easily observe the degree of association between nationalities and identify which groups exhibit stronger residential ties, offering a clearer picture of the underlying shared residence dynamics in Vienna.

\begin{figure}[bt!]
\centering
\includegraphics[width=\textwidth]{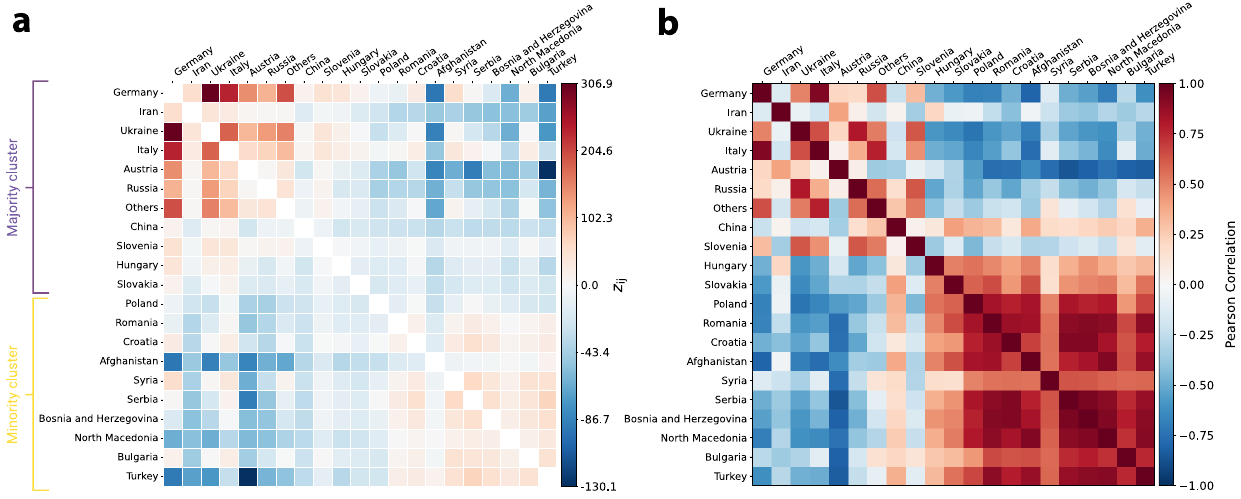}
\caption[Significant shared residence links and strength of co-residence.]{\textbf{Significant shared residence links and strength of co-residence.} \textbf{a} \textit{z}-scores matrix. \textbf{b} Pearson Correlation matrix. Nationalities are ordered according to the Infomap clustering results. Co-residence clusters are marked with the same colour coding (without Bonferroni). The exact values for both matrices are available in the GitHub repository \cite{github}.}
\label{fig:figs6}
\end{figure}

\subsection{Robustness assessment} \label{sec:sanity}
In this subsection, we assess the robustness of our results to ensure the reliability and stability of the identified co-residence patterns. By performing a series of sensitivity analyses and alternative tests, we evaluate whether the observed relationships hold under different assumptions and conditions. This robustness assessment helps confirm that the patterns we uncover are not driven by random noise or methodological artefacts but reflect meaningful underlying dynamics in the data.

\subsubsection{Bonferroni correction} \label{sec:bonferrini}
To ensure the statistical reliability of our results, we apply the Bonferroni correction, a widely used method for addressing the problem of multiple comparisons \cite{abdi2007bonferroni}. By adjusting the significance threshold, this correction minimises the risk of false positives, allowing us to robustly identify meaningful shared residence patterns while accounting for the increased likelihood of chance findings when testing numerous hypotheses. 

Using the Bonferroni correction, we establish a criterion for determining the significance of shared residence links between nationalities. Specifically, a link is considered statistically significant if the probability of observing its total \textit{z}-score (as defined in Eq. (\ref{eq:zscore}) of the main document) is less than 0.05/C, where $C=21$ is the number of diverse nationalities residing in Vienna. We divide by 21 because, for each nationality, we test its co-residence associations with the remaining nationality groups. This threshold ensures that the probability of incorrectly identifying a link as significant is appropriately adjusted for the multiple comparisons. Given that the total \textit{z}-score is calculated as a sum over many independent variables, we can approximate its expected distribution using a normal distribution. This normal distribution is centred at zero and has a standard deviation of $\sqrt{D}$, where $D=23$ is the number of districts in Vienna. This methodology is based on the one used in \cite{karimi2015mapping}, where the authors identify significant bilateral information interests between countries based on the activity of Wikipedia editors.

Using these parameters, we compute the significance threshold for the total \textit{z}-score as $t=2.82\sqrt{D}$, where 2.82 is derived from the condition that $P(z>2.82) = 0.05/C$, and \textit{P} is the standard Gaussian distribution (with zero average and unit variance). Consequently, if the observed total \textit{z}-score for a pair of nationalities \textit{i} and \textit{j} exceeds this threshold, we establish a significant link between these countries with weight $\Tilde{z}_{ij}$ according to:
\begin{equation}
\tilde{z}_{i j}=\left\{\begin{array}{cc}
z_{i j}-t & \text { if } z_{i j}>t\text{, and} \\
0 & \text { if } z_{i j} \leqslant t.
\end{array}\right.
\end{equation}

\begin{figure*}[bt!]
\centering
\includegraphics[width=\textwidth]{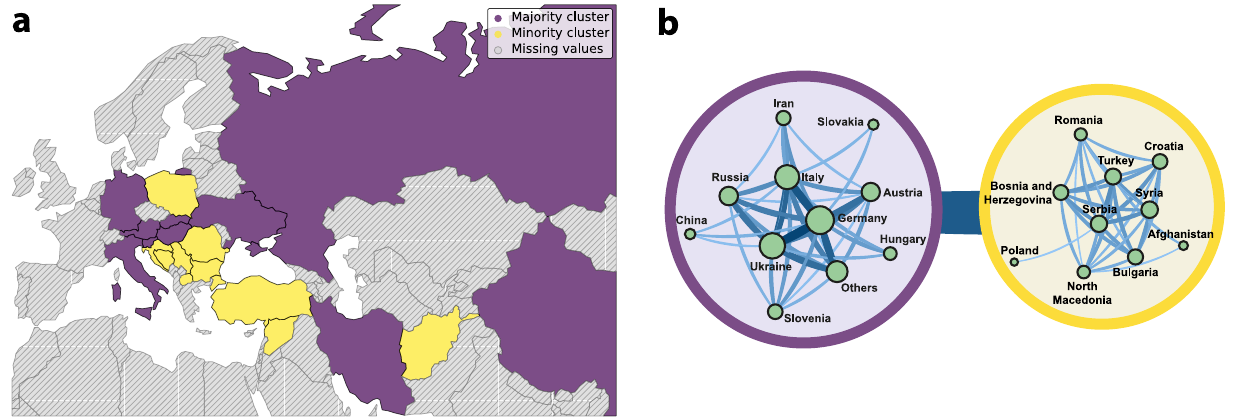}
\caption[Clustering nationalities based on district co-residence patterns without Bonferroni correction.]{\textbf{Clustering nationalities based on district co-residence patterns without Bonferroni correction.} \textbf{a} World map of district co-residence patterns. Countries that belong to the same cluster have the same colour. \textbf{b} World network of district co-residence patterns. The size of the nodes represents the total $\textit{z}$-score of the clusters and countries. The links represent the connections between nodes obtained from the cluster analysis with Infomap \cite{mapequation2023software}, the thicker the line, the stronger the connection.}
\label{fig:figs8}
\end{figure*}

Community detection results applying the Bonferroni correction before Infomap clustering \cite{mapequation2023software} are shown in Fig. \ref{fig:fig2}. Removing this correction does not yield any qualitative changes in the results (Supplementary Fig. \ref{fig:figs8}). The primary impact is the integration of the four countries (China, Slovakia, Afghanistan, and Poland) to the rest. This results in these nations being among the least connected within their cluster, suggesting that other factors may account for their inclusion.
The model preserves the average population size of each nationality while randomising the districts in which they reside. By comparing the results of the interest model with the empirical data, we can identify significant shared residence links among nationalities, highlighting patterns that deviate from random expectations. Besides, this rigorous filtering method helps isolate the most meaningful patterns of co-residence while accounting for the broader context of multiple testing.

\subsubsection{Pearson correlation coefficient}

The Pearson correlation coefficient, denoted as $r$, measures the strength and direction of a linear relationship between two continuous variables. It ranges from $-1$ to $+1$, where $r = +1$ indicates a perfect positive linear relationship, $r = -1$ denotes a perfect negative linear relationship, and $r = 0$ suggests no linear association between the variables \cite{pearson1895vii}. Interpretation of \textit{r} typically follows Cohen’s guidelines, which suggest that $|r| < 0.3$ represents a weak relationship, $0.3 \leq |r| < 0.5$ indicates a moderate relationship, and $|r| \geq 0.5$ suggests a strong relationship \cite{cohen2013statistical}. However, these cutoffs are context-dependent, and scientific judgment should consider the practical significance of the relationship in the specific field of study.

To compute $r$, we first assume two paired variables, $X$ and $Y$, with $n$ paired observations $(X_i, Y_i)$ for $i = 1, 2, \dots, n$. The formula for $r$ is given by:
\begin{equation}
    r = \frac{\sum_{i=1}^{n} (X_i - \bar{X})(Y_i - \bar{Y})}{\sqrt{\sum_{i=1}^{n} (X_i - \bar{X})^2} \sqrt{\sum_{i=1}^{n} (Y_i - \bar{Y})^2}},
    \label{eq:Pearsoncorrelation}
\end{equation}
where $\bar{X}$ and $\bar{Y}$ are the sample means of $X$ and $Y$, respectively. This equation standardises the covariation between $X$ and $Y$ by dividing by their product of standard deviations, yielding a dimensionless value that facilitates comparison across datasets and units \cite{fisher1970statistical}.

To quantify the uncertainty of $r$, we calculate its standard error, $\sigma_r$, given by:
\begin{equation}
    \sigma_r = \sqrt{\frac{1 - r^2}{n - 2}},
    \label{eq:correlationerror}
\end{equation}
where $n$ is the sample size. The standard error measures the extent to which the observed $r$ might vary due to sampling variability, providing an estimate of its precision.

\subsubsection{Measuring strength of co-residence} 
We here evaluate the strength of co-residence using Pearson correlations between countries based on their population fractions (Supplementary Fig. \ref{fig:figs6}\textbf{b}). This metric provides an additional layer of validation, confirming that the shared residence patterns derived from our primary analysis are consistent with the Infomap clustering results. Specifically, positive correlations indicate that countries with stronger district-level shared residence patterns tend to cluster together, whereas negative correlations highlight divisions between clusters. Hungary and Slovakia stand as exceptions, exhibiting an unexpected pattern of co-residence despite belonging to different clusters. 

This consistency across methods strengthens the robustness of our findings, as it demonstrates that the clustering outcomes are not arbitrary but rather reflect coherent and meaningful relationships between nationalities. Based on this initial robustness check, we conclude that the clustering results remain qualitatively stable and unaffected by methodological artefacts.

\subsubsection{District-level population fraction deviations from city-wide expectations in Vienna} 
Understanding how district-level population fractions deviate from city-wide averages provides a valuable perspective on residential patterns and segregation. By examining these deviations, we aim to identify which districts house disproportionately higher or lower populations of specific nationalities relative to their overall representation in Vienna. 

In this subsection, we assess how each district’s population fraction for a specific nationality compares to that nationality’s city-wide population fraction. To achieve this, we analyse the population fraction of each country relative to the total population in each district (Supplementary Fig. \ref{fig:figs4}). This calculation highlights which districts exceed or fall short of the expected population fractions, uncovering patterns of over- or under-representation.

\begin{figure}[bt!]
\centering
\includegraphics[width=\textwidth]{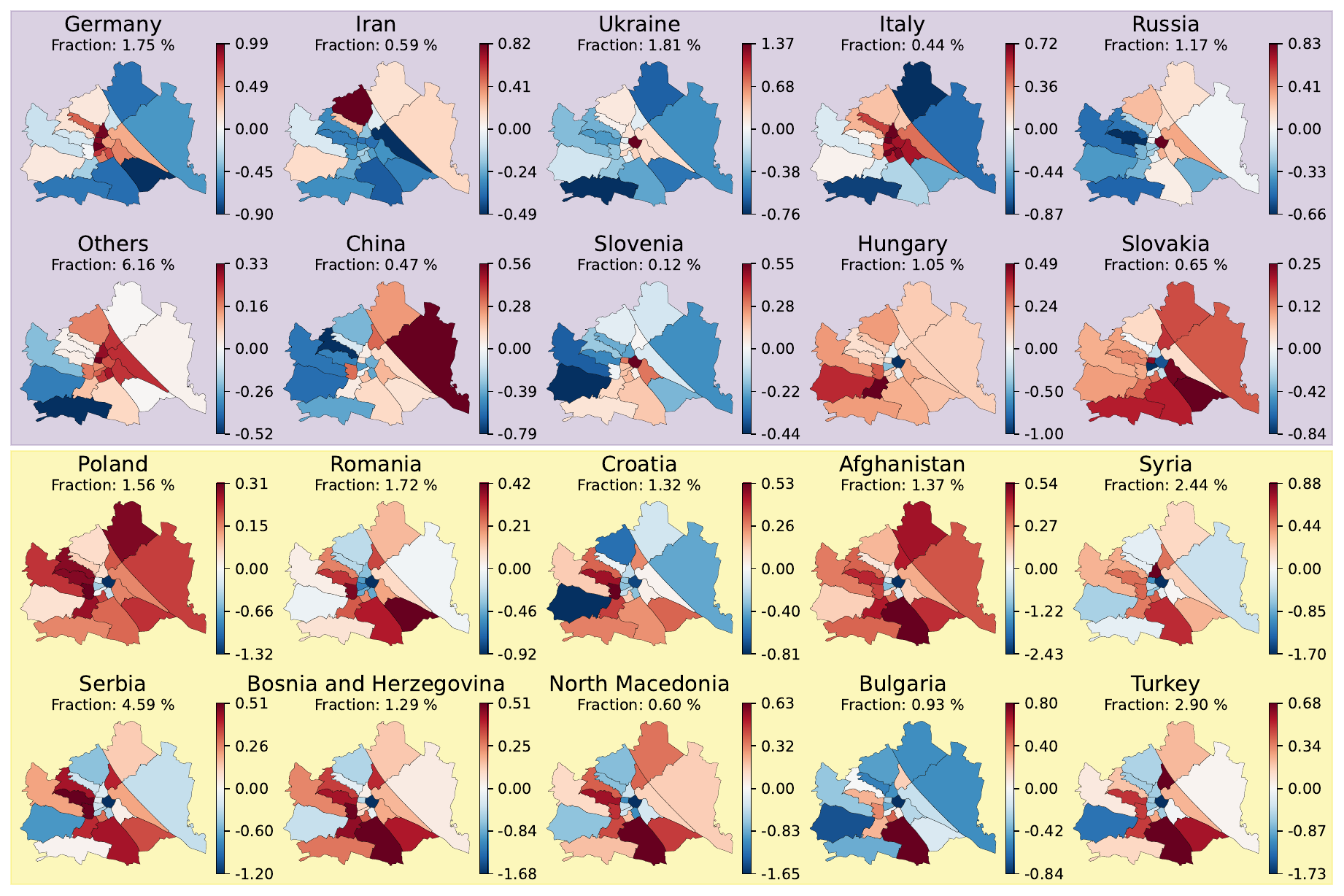}
\caption[District-level population fraction deviations from city-wide expectations in Vienna.]{\textbf{District-level population fraction deviations from city-wide expectations in Vienna.} Nationalities are arranged based on the Infomap clustering results. Co-residence clusters are shown with the same colour coding (without Bonferroni). Deviations are represented in log-scale for improved interpretability.}
\label{fig:figs7}
\end{figure}

Countries in the minority cluster exhibit similar districts in which their populations are overrepresented relative to citywide expectations (Supplementary Fig. \ref{fig:figs7}). Likewise, the majority cluster also demonstrates shared districts of higher-than-expected representation. These patterns reinforce earlier clustering results and provide a localised view of how residential segregation manifests at the district level.

Through this robustness analysis, we confirm that deviations in district-level population fractions align with the clustering patterns identified earlier. This further supports the validity of our findings and highlights the nuanced residential behaviours shaping Vienna's urban landscape.

\subsubsection{Tuning Infomap's Markov time parameter} 

The Infomap search algorithm includes a Markov time parameter that tunes the cost of a random walker's transition between modules. The default value is $T_{\mathrm{mk}}=1$. For higher values, the algorithm yields larger clusters; conversely, for lower values, it yields smaller clusters. To assess cluster stability, we analyzed the Bonferroni-corrected network varying the Markov time parameter, and found that the reported partition is unchanged over a broad interval, $T_{\mathrm{mk}} \in [0.7,\,3.8]$.

Outside this range, the solution changes in an interpretable way:
\begin{itemize}
    \item $T_{\mathrm{mk}} > 3.9$: all countries merge into a single cluster.
    \item $T_{\mathrm{mk}} = 0.6$: Hungary separates into its own cluster (remaining most strongly connected to the purple cluster).
    \item $T_{\mathrm{mk}} = 0.5$: the purple cluster fragments, while the yellow cluster remains intact.
    \item $T_{\mathrm{mk}} = 0.3$: all countries split into singleton clusters.
\end{itemize}

\begin{figure*}[bt!]
\centering
\includegraphics[width=\textwidth]{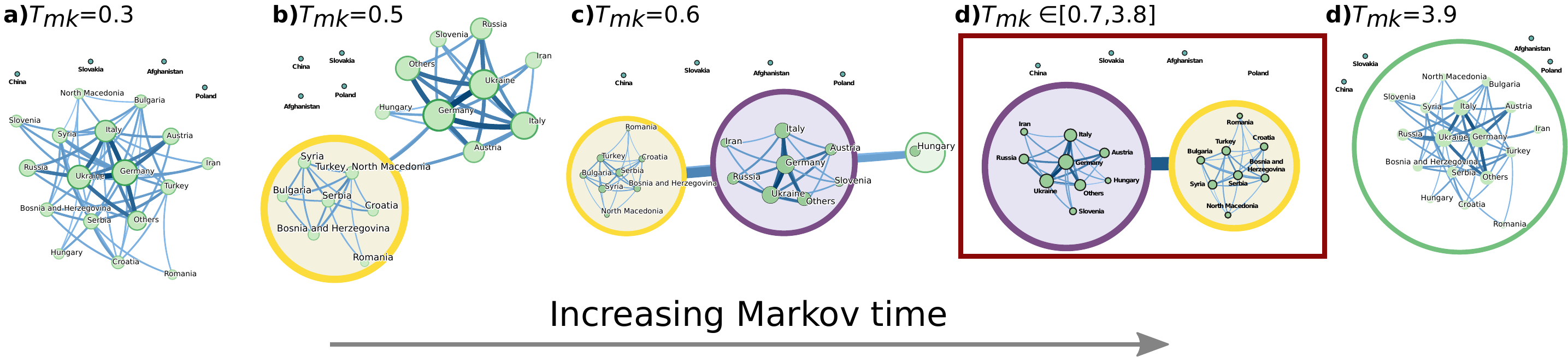}
\caption[Stability of Infomap clusters under varying Markov time.]{\textbf{Robustness of clusters under varying Markov time.} Clusters of the country–country interaction network (nodes = nationalities; edge weights proportional to co-residence–based interaction strength) clustered with Infomap at increasing Markov time $T_{\mathrm{mk}}$. At very small $T_{\mathrm{mk}}$ the partition fragments into (near-)singleton communities (a, $T_{\mathrm{mk}}=0.3$); intermediate values produce the same two-community structure reported in the main text (d, $T_{\mathrm{mk}}\in[0.7,3.8]$, red box), with only minor changes close to the boundary (b–c, $T_{\mathrm{mk}}=0.5$–0.6). For sufficiently large $T_{\mathrm{mk}}$, all countries merge into a single community (e, $T_{\mathrm{mk}}=3.9$).}
\label{fig:markov_time}
\end{figure*}

We also applied a variant of Infomap with variable Markov time \cite{PhysRevE.93.032309} to recover communities at different effective scales; this procedure yields the same clusters as those reported in the manuscript (Fig.~\ref{fig:markov_time}d).

\section{Determinants of co-residence clusters}\label{sec:determinants}
We introduce this section to better understand the factors that can contribute to the formation of the two distinct clusters identified in the previous analysis. Specifically, we examine how national homophily, neighbourhood income, and neighbourhood diversity shape Vienna residents' residence tendencies. To quantify the impact of economic factors on district-level shared residence patterns, we incorporate additional publicly accessible datasets to thoroughly analyse our clustering results. Our investigation reveals that these dimensions significantly influence the formation and segregation of clusters within Vienna's population.

Economic disparities create conditions for segregation, as wealthier residents can afford to live in more affluent districts. At the same time, geographical and national homophily reinforce these patterns by encouraging residents to settle in areas with others who share similar backgrounds. These factors serve as both pull and push factors influencing citizens' residential decisions. Understanding the underlying causes of residential segregation is crucial for developing policies that foster social cohesion and integration. Addressing both economic and homophily factors can help design interventions that promote more diverse and inclusive neighbourhoods.

\subsection{Measuring national homophily} 

We examine national homophily as a key factor influencing residential clustering within Vienna. To quantify this tendency, we compute the Dissimilarity Index for each nationality, as shown in Supplementary Table \ref{tab:tabs4}. This index measures the evenness of a given nationality’s distribution across Vienna's districts relative to the overall population, providing a quantitative assessment of segregation levels.

\subsection{Measuring neighbourhood income}

We investigate the role of neighbourhood income in shaping residential tendencies and clustering patterns in Vienna. Income-related factors play a significant role in determining where social groups choose to live. To measure the impact of neighbourhood income on district-level shared residence, we use district-level average net income data and examine how income disparities shape the formation of residential clusters. Additionally, in this Supplementary Information, we also examine how rental prices across Vienna's districts influence the formation of these clusters. By integrating these economic variables into our analysis, we aim to better understand how economic factors contribute to the spatial segregation of different nationality groups in Vienna. 

In relation to the results presented in the main manuscript, Supplementary Table \ref{tab:tabs4} provides the exact values for the Income-Population fraction correlation. To further extend our analysis, we also compute the Pearson correlation between the percentage of foreigners and the average net income per district, obtaining a value of \textit{r}[\% foreigners, Income] = -0.80 $\pm$ 0.13 (see Supplementary Fig. \ref{fig:figs11}\textbf{b}). This strong negative correlation indicates that foreigners are more likely to reside in districts with lower average net income. Additionally, we investigate the Pearson correlation between the percentage of foreigners and the average rental price per $m^2$, which yields a value of \textit{r}[\% foreigners, Rental] = -0.43 $\pm$ 0.20 (Supplementary Fig. \ref{fig:figs11}\textbf{c}). This negative correlation suggests that foreigners are more likely to reside in districts with lower rental prices.

We explore alternative perspectives on neighbourhood income by stratifying individuals using two distinct methods for both datasets. First, we present an estimated neighbourhood income box plot for each nationality, providing a visual representation of income distribution. Second, we partition individuals into four equal-sized quartiles based on either the median household income within their residential areas or the median household rental price \cite{moro2021mobility}. These approaches offer a more granular understanding of the relationship between neighbourhood income and residential clustering.

\subsubsection{Neighbourhood average net income} 

\paragraph{Box plot perspective.}
To estimate the neighbourhood income distribution for each nationality residing in Vienna, we approximate it at the district level by assuming that all inhabitants of a given district earn the average net income of that district (Supplementary Fig. \ref{fig:figs2}\textbf{a}). For each nationality, we construct a box plot representing the average income distribution. Each box plot highlights the median income of the population of the given nationality (green horizontal line) alongside the corresponding quartiles (Supplementary Fig. \ref{fig:figs9}\textbf{a}).

To compute these box plots, we employed the following methodology: for each nationality, we first sorted the districts by their population fraction (Supplementary Fig. \ref{fig:figs3}), starting with the district with the lowest fraction of the nationality's population and proceeding to the one with the highest fraction. We then computed the cumulative sum of these fractions, starting with the smallest and adding each district's fraction in ascending order. When the cumulative sum exceeded the target values for the quartiles or the median (i.e., the 25\%, 50\%, and 75\% thresholds of the nationality's population in Vienna), we interpolated between the two bounding districts using linear regression. This interpolation allowed us to pinpoint the district's average net income at these threshold points.

By applying this systematic approach across all nationalities, we derived a robust representation of the neighbourhood income distributions. This method provides an intuitive, yet quantitative, understanding of how the population of each nationality is distributed across Vienna's socio-economic landscape. It also offers a reliable approximation for the variation in neighbourhood income experienced by different nationalities residing in the city.

There are income disparities among residents within each nationality (Supplementary Fig. \ref{fig:figs9}\textbf{a}). Furthermore, by ranking countries by median income from highest to lowest, we highlight the stark inequalities across populations of different nationalities residing in Vienna. This approach enables the development of a metric that captures income inequalities on two distinct scales: within each nationality and between different nationalities. This economic division reinforces clustering patterns, as the two groups with similar co-residence tendencies are completely segregated with respect to neighbourhood income distributions.

The analysis indicates that the countries in the majority cluster reside in districts with higher average net income than those in the minority cluster. Thus, residential segregation among nationalities may be partly driven by economic factors. Specifically, individuals from wealthier countries are more likely to reside in affluent districts, whereas those from less economically advantaged nations tend to live in lower-income areas. To substantiate these findings, we also examined the extent to which residents of various nationalities living in Vienna represent the socio-economic profiles of their countries of origin (Supplementary Fig. \ref{fig:figs9}\textbf{a}). There is a correlation is observed between the median income in Vienna and the median income in the country of origin. This indicates that citizens residing in Vienna generally reflect the socio-economic characteristics of their home countries. However, exceptions include foreigners from Iran and Ukraine, whose median incomes in Vienna are uncorrelated with those of their home countries. This suggests that individuals from these nations are likely to be better qualified or more economically advantaged than the average citizen in their countries of origin.

\paragraph{Quartile-based perspective.}

From the quartile-based perspective, we aim to provide a complementary view of neighbourhood income distribution for each nationality residing in Vienna. While the box plot perspective captures the distribution of income levels across districts for each nationality, the quartile-based approach instead categorises districts into four quartiles based on their average net income in Vienna. This methodology allows us to summarise how national populations are distributed across districts of varying economic statuses in a simple, visually intuitive format.

To construct the quartile-based bar plots, we first divide all Vienna districts into four quartiles according to their average net income, with the first quartile representing districts with the lowest incomes and the fourth quartile representing districts with the highest incomes (Supplementary Fig. \ref{fig:figs2}\textbf{b}). Next, for each nationality, we calculate the fraction of its population residing in each district relative to the total population of that nationality in Vienna (Supplementary Fig. \ref{fig:figs3}). For each quartile, we sum the population fractions of all districts within that quartile for a given nationality. This process yields a set of four values for each nationality, representing the proportion of its population residing in districts in each income quartile (Supplementary Fig. \ref{fig:figs9}\textbf{b}). By comparing these stacked bar plots by nationality, we can observe how populations across different countries are distributed across districts with varying levels of affluence in Vienna. Therefore, we rank nationalities by cumulative population fractions up to the second quartile, as this metric provides an analogous representation of the estimated median. This ranking enables us to quantify the proportion of each nationality residing in districts in the lower half of Vienna’s income distribution. By doing so, we reveal patterns of inequality both within and between nationalities.

Once again, clear disparities emerge between the two identified clusters of co-residence tendencies: the majority and minority clusters. Nationalities belonging to the majority cluster tend to have significantly lower cumulative fractions in the first and second quartiles, indicating their populations are concentrated in wealthier districts of Vienna. Conversely, minority nationalities exhibit higher cumulative fractions in the lower-income quartiles, indicating a greater prevalence in economically disadvantaged areas. The ranking confirms the economic segregation hypothesis. The clustering of nationalities into distinct shared residence groups appears strongly influenced by income-related factors, with wealthier districts predominantly occupied by populations from wealthier countries. Meanwhile, populations from less affluent countries are more likely to reside in lower-income districts.

This approach provides insights similar to the box plot perspective but emphasises the relative representation of each nationality within different income brackets across the city, offering a clearer view of the extent to which various populations are concentrated in wealthier or less affluent areas. This visualisation also allows us to compare disparities in economic integration and resource access across nationalities in a straightforward, standardised format.

\subsubsection{Neighbourhood Rental price} \label{sec:rental}

\begin{figure*}[bt!]
\centering
\includegraphics[width=\textwidth]{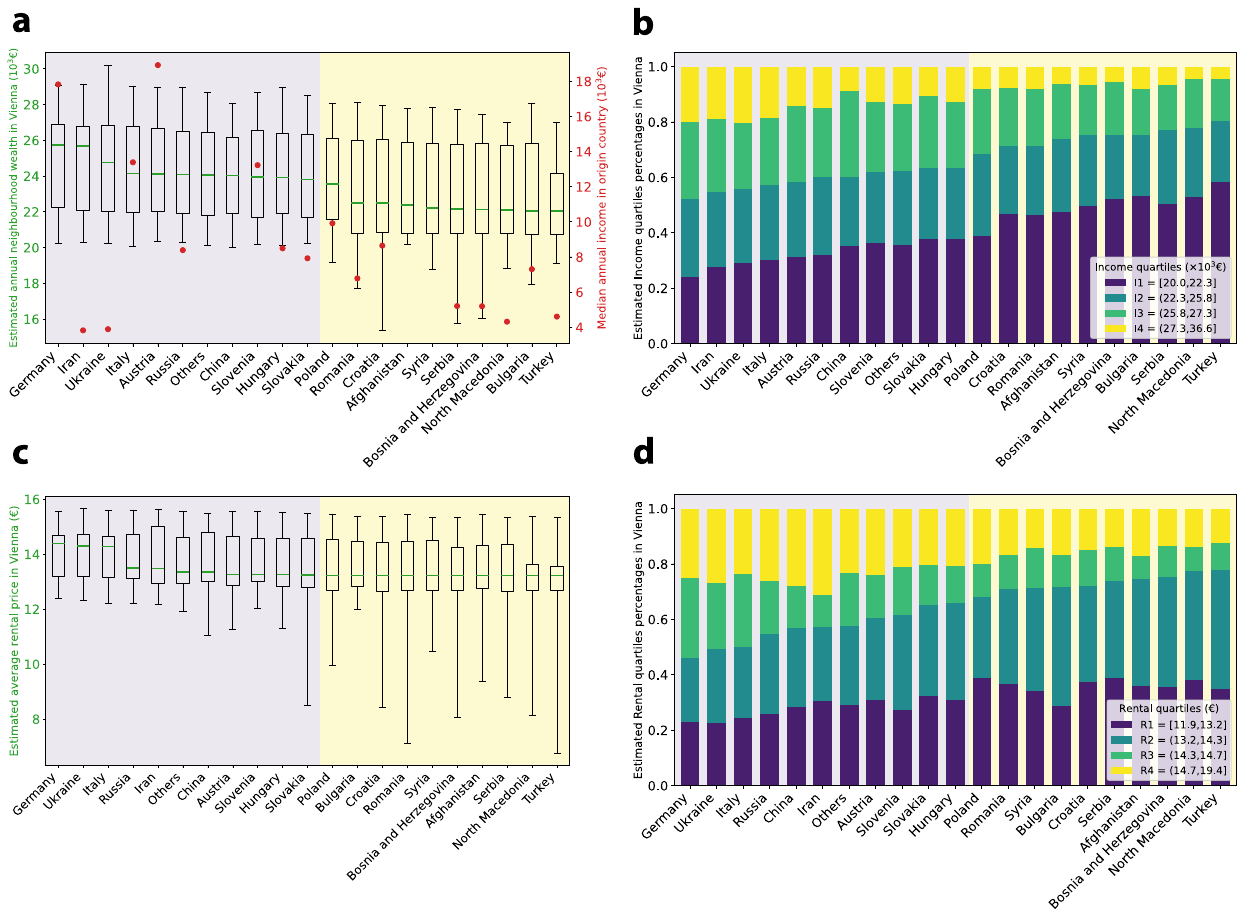}
\caption[Neighbourhood income distributions.]{\textbf{Neighbourhood income distributions.} \textbf{a} Estimated annual neighbourhood average income distributions of residents from each country living in Vienna. Nationalities are ranked from highest to lowest by estimated median (green horizontal lines). Median annual incomes in origin countries are also included (red dots). The dataset lacks information related to China, Afghanistan, and Syria. \textbf{b} Estimated Income quartiles percentages of each country living in Vienna. Nationalities are sorted from lowest to highest based on the cumulative population fractions up to the second quartile. \textbf{c} Estimated average rental price distributions of residents from each country living in Vienna. Nationalities are ranked from highest to lowest by estimated median (green horizontal lines). \textbf{d} Estimated Rental quartiles percentages of each country living in Vienna. Nationalities are sorted from lowest to highest based on the cumulative population fractions up to the second quartile.}
\label{fig:figs9}
\end{figure*}

\paragraph{Box plot perspective.}
To estimate the distribution of neighbourhood rental prices for each nationality residing in Vienna, we use a similar approach as outlined for neighbourhood income, but now assuming that all inhabitants of a given district pay the
average rental price of that district (Supplementary Fig. \ref{fig:figs2}\textbf{c}). For each nationality, we construct a box plot showing the distribution of average rental prices experienced by its population. The box plot highlights the median rental price (green horizontal line) and the corresponding quartiles for each nationality (Supplementary Fig. \ref{fig:figs9}\textbf{c}).

The methodology for constructing these box plots mirrors the approach used for income distributions. Specifically, districts are first sorted by their population fraction for each nationality, and a cumulative sum is performed to determine the quartiles and median. The interpolation process allows us to identify the corresponding rental price values at these thresholds. This ensures consistency in methodology and provides a comparable representation of neighbourhood income, now assessed through rental prices.

Nationalities belonging to the majority cluster predominantly reside in districts with higher average rental prices, while those in the minority cluster are concentrated in districts with lower rental prices (Supplementary Fig. \ref{fig:figs9}\textbf{c}). By ranking nationalities from highest to lowest median rental price, we observe the same segregation into majority and minority clusters, reinforcing the hypothesis that economic factors, be it net income or rental price, play a significant role in shaping residential clustering. The patterns derived from rental price data closely parallel those observed in the income-based analysis. Both studies highlight a significant economic division between the majority and minority cluster.

\paragraph{Quartile-based perspective.}
The quartile-based perspective for rental prices provides a complementary view of the neighbourhood rental price distribution for each nationality residing in Vienna, analogous to the approach used for income. Here, districts are categorised into four quartiles based on their average rental price (Supplementary Fig. \ref{fig:figs2}\textbf{d}), and the population fraction of each nationality residing in districts within each quartile is summed. This methodology enables us to summarise how national populations are distributed across districts with varying levels of rental affordability (Supplementary Fig. \ref{fig:figs9}\textbf{d}).

The results closely mirror those observed in the income analysis: nationalities in the majority cluster are concentrated in districts with higher rental prices, while those in the minority cluster tend to reside in districts with lower rental prices. By ranking nationalities based on their cumulative population fractions up to the second quartile, we once again observe disparities both within and between nationalities. These disparities underscore the economic segregation between the two clusters identified in the shared residence analysis, with wealthier populations occupying higher-rent districts and less affluent populations residing in more affordable areas. The quartile-based perspective for rental prices corroborates the findings from the income analysis, further highlighting the critical role of economic factors in shaping residential segregation in Vienna.

\subsubsection{Relationship between Income and Rental price: A clustering approach} 
\begin{figure*}[bt!]
\centering
\includegraphics[width=\textwidth]{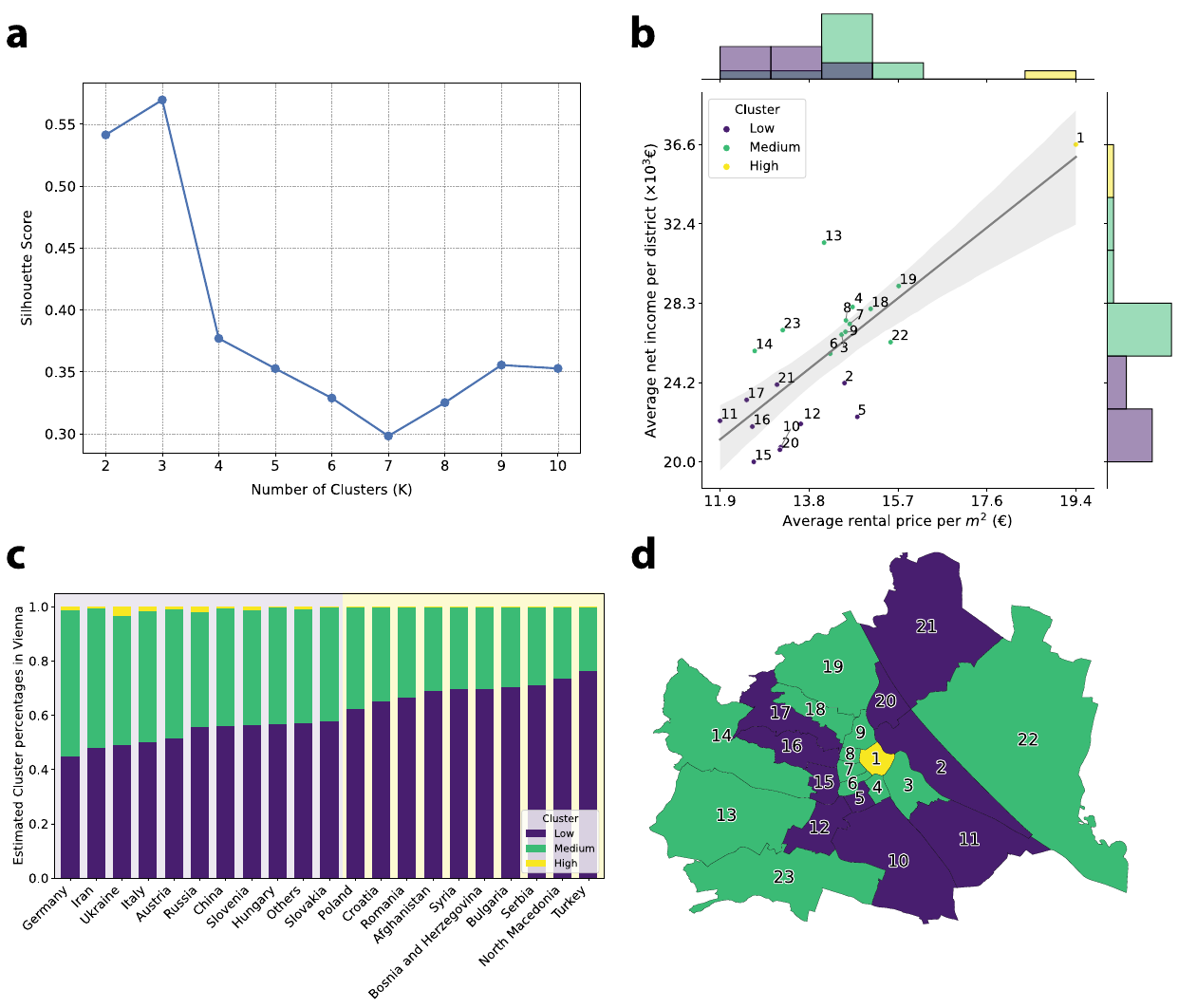}
\caption[Relationship between Income and Rental price: A clustering approach]{\textbf{Relationship between Income and Rental price: A clustering approach} \textbf{a} Silhouette score used to determine the optimal number of clusters (K). \textbf{b} Scatter plot showing the clustering of average rental prices versus average net income in Vienna districts, including the regression line and histograms of cluster distributions. \textbf{c} Estimated percentage of the population of each nationality residing in districts belonging to the low, medium, and high SES clusters. Nationalities are ordered from lowest to highest by cumulative population fraction in the low-SES cluster. \textbf{d} Geographical map of Vienna displaying the spatial distribution of the identified clusters.}
\label{fig:figs10}
\end{figure*}

To explore the interplay between neighbourhood income and rental prices in Vienna, we employ a K-means clustering approach. K-means is a widely used clustering algorithm that partitions data points into K distinct clusters by minimising the variance within each cluster \cite{macqueen1967some}. In this analysis, we use the Silhouette score technique to determine the optimal number of clusters, identifying K=3 as the best fit (Supplementary Fig. \ref{fig:figs10}\textbf{a}). The resulting clusters represent districts with low, medium, and high socio-economic status (SES) based on their average net income and average rental prices.

The clustering process begins by plotting the relationship between average rental price and average net income for the 23 districts of Vienna. A regression line is fitted to assess the correlation between these two variables (see Supplementary Fig. \ref{fig:figs10}\textbf{b}). The analysis reveals a positive correlation, indicating that districts with higher average incomes tend to have higher rental prices. Using the identified clusters, we visualise the distribution of districts in a scatter plot, where the clusters are colour-coded as low, medium, and high SES. Histograms of income and rental price distributions within each cluster are also provided, offering additional insights into their distinct characteristics.

Building on this clustering approach, we quantify the residential patterns of different nationalities by summing the fraction of each nationality’s population residing in districts belonging to the low, medium, or high SES clusters. This yields a bar plot for each nationality, with cluster percentages summing to 1 (Supplementary Fig. \ref{fig:figs10}\textbf{c}). By ranking these bars based on the cumulative population fraction in the low cluster, we observe stark patterns of segregation. Nationalities in the majority cluster, as identified in the Infomap co-residence analysis, predominantly reside in districts within the high or medium SES clusters. Conversely, nationalities in the minority cluster are disproportionately represented in districts categorised as low SES cluster. This segregation underscores the economic divide between the two groups, reinforcing the hypothesis that socio-economic factors drive residential clustering in Vienna.

The geographical distribution of the three clusters is mapped across Vienna (see Supplementary Fig. \ref{fig:figs10}\textbf{d}).
The spatial arrangement reveals a clear pattern: high- and medium-SES districts are concentrated in two main areas—a dense core surrounding the first district and a few suburban pockets on the outskirts. In contrast, low-SES districts form a contiguous ring encircling the core of more affluent neighbourhoods. This map underscores the role of SES in shaping geographical segregation in Vienna, as districts with similar SES levels tend to cluster into connected regions.

The K-means clustering approach reveals a clear relationship between income and rental price distributions in Vienna’s districts. The clustering results align with and further validate the segregation observed in the shared residence analysis. Populations from wealthier countries are concentrated in districts with higher incomes and rents, whereas populations from less affluent countries reside in economically disadvantaged areas. This analysis highlights the intricate link between income, rental affordability, and residential segregation in Vienna.

\subsection{Measuring neighbourhood diversity} \label{sec:diversity}

We investigate neighbourhood diversity as a key factor influencing residential clustering within Vienna. To quantify this tendency, we compute the Simpson Index for each district (Supplementary Table \ref{tab:tabs3}). This index expresses the probability that two randomly selected individuals from the district/city have different nationalities, offering a quantitative assessment of concentration levels. 

\begin{figure*}[bt!]
\centering
\includegraphics[width=\textwidth]{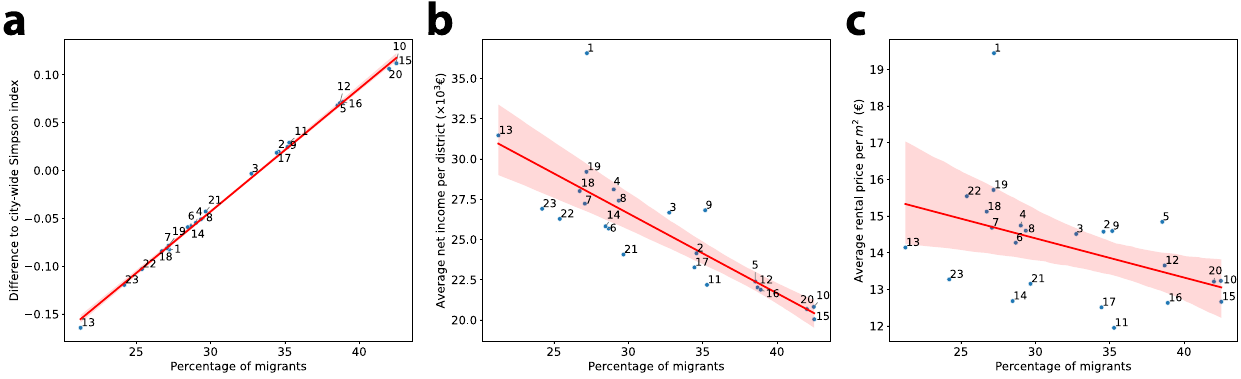}
\caption[Correlations between key variables]{\textbf{Correlations between key variables.} \textbf{a} Correlation between neighbourhood diversity and the percentage of foreigners in Vienna's districts. \textbf{b} Correlation between average net income and the percentage of foreigners in Vienna's districts. \textbf{c} Correlation between average rental price and the percentage of foreigners in Vienna's districts.}
\label{fig:figs11}
\end{figure*}

The Supplementary Table \ref{tab:tabs4} provides the exact values for the Diversity-Population fraction correlation. To further extend our analysis, we also compute the Pearson correlation between the percentage of foreigners and the difference to the city-wide Simpson index per district, obtaining a value of \textit{r}[\% foreigners, Diversity]=0.99 $\pm$ 0.01 (Supplementary Fig. \ref{fig:figs11}\textbf{a}). This strong positive correlation indicates that foreigners are more likely to reside in districts with higher diversity.

\begin{table}[bt!]
    \centering
    \caption[Neighbourhood diversity indices.]{\textbf{Neighbourhood diversity indices.} The Simpson and Entropy indices for each district. Districts are ranked from highest to lowest by the estimated Simpson index for Vienna districts.}
    \begin{tabular}{c|c|c}
    \begin{tabular}{c}
    \textbf{District}
    \end{tabular} & \begin{tabular}{c}
    \textbf{Simpson} \\ \textbf{index}
    \end{tabular} & \begin{tabular}{c}
    \textbf{Entropy} \\ \textbf{index}
    \end{tabular} \\
        \hline
        \hline
        10 & 0.653 & 1.801 \\
        15 & 0.653 & 1.804 \\
        20 & 0.647 & 1.782 \\
        16 & 0.612 & 1.682 \\
        12 & 0.611 & 1.698 \\
        5 & 0.609 & 1.691 \\
        11 & 0.570 & 1.576 \\
        9 & 0.565 & 1.519 \\
        2 & 0.560 & 1.548 \\
        17 & 0.559 & 1.550 \\
        3 & 0.537 & 1.496 \\
        21 & 0.498 & 1.403 \\
        8 & 0.490 & 1.328 \\
        4 & 0.486 & 1.330 \\
        6 & 0.482 & 1.339 \\
        14 & 0.481 & 1.361 \\
        19 & 0.462 & 1.294 \\
        7 & 0.459 & 1.254 \\
        1 & 0.458 & 1.188 \\
        18 & 0.456 & 1.284 \\
        22 & 0.438 & 1.251 \\
        23 & 0.421 & 1.215 \\        
        13 & 0.376 & 1.090 \\
    \end{tabular}
    \label{tab:tabs3}
\end{table}

\begin{table}[bt!]
    \centering
    \caption[Country-specific socio-economic indicators.]{\textbf{Country-specific socio-economic indicators.} Correlation of income and diversity with population fraction, alongside the Dissimilarity and Gini indices for each country. Nationalities are ordered according to the Infomap clustering results.}
    \begin{tabular}{c|c|c|c|c}
    \begin{tabular}{c}
    \textbf{Country}
    \end{tabular} & \begin{tabular}{c}
    \textbf{Income-Population} \\ \textbf{fraction correlation}
    \end{tabular} & \begin{tabular}{c}
    \textbf{Diversity-Population} \\ \textbf{fraction correlation}
    \end{tabular} & \begin{tabular}{c}
    \textbf{Dissimilarity} \\ \textbf{index}
    \end{tabular} & \begin{tabular}{c}
    \textbf{Gini} \\ \textbf{index}
    \end{tabular} \\
        \hline
        \hline
        Germany & 0.388 & -0.204 & 0.237 & 0.291 \\
        Iran & 0.260 & -0.401 & 0.142 & 0.172 \\
        Ukraine & 0.700 & -0.213 & 0.187 & 0.260 \\
        Italy & 0.352 & -0.035 & 0.214 & 0.245 \\
        Austria & 0.800 & -0.999 & 0.130 & 0.051 \\
        Russia & 0.501 & -0.073 & 0.130 & 0.196 \\
        Others & 0.001 & 0.358 & 0.079 & 0.108 \\
        China & -0.406 & 0.335 & 0.137 & 0.191 \\
        Slovenia & 0.271 & 0.071 & 0.086 & 0.134 \\
        Hungary & -0.400 & 0.162 & 0.063 & 0.130 \\
        Slovakia & -0.517 & 0.307 & 0.069 & 0.140 \\
        Poland & -0.794 & 0.567 & 0.091 & 0.188 \\
        Romania & -0.826 & 0.755 & 0.124 & 0.203 \\
        Croatia & -0.808 & 0.746 & 0.143 & 0.204 \\
        Afghanistan & -0.791 & 0.617 & 0.161 & 0.290 \\
        Syria & -0.633 & 0.761 & 0.199 & 0.307 \\
        Serbia & -0.880 & 0.857 & 0.180 & 0.252 \\
        Bosnia and Herzegovina & -0.883 & 0.807 & 0.181 & 0.267 \\
        North Macedonia & -0.834 & 0.738 & 0.199 & 0.316 \\
        Bulgaria & -0.722 & 0.803 & 0.211 & 0.238 \\
        Turkey & -0.842 & 0.820 & 0.241 & 0.343 \\
    \end{tabular}
    \label{tab:tabs4}
\end{table}

\section{From residential segregation to neighbourhood segregation}

{
Although the core objective is to study residential segregation, it is relevant to contextualise our units of observation, neighbourhoods, as spatially embedded within neighbouring districts and administrative boundaries. The districts were observed as bounded containers intended to encompass opportunities for integration and encounters among national groups. Yet residents of one district may have frequent interactions with people from nearby areas, even if those individuals belong to a different district. This is particularly relevant when the border between one district and an adjacent one is a residential street, as is the case between Josefstadt, the 8th district, and Neubau, the 7th, delineated by Lerchenfelder Str.
}

{
To assess the degree of segregation in terms of neighbourhoods, we first analyse whether any two neighbourhoods are adjacent, meaning they share a border of some length, and we compare the distribution of the population under a scenario in which each person moves randomly to an adjacent district. The principle is that the distribution of people from two countries might show quite different patterns, but if, after moving to an adjacent neighbourhood, perhaps as part of their journeys to satisfy basic needs, those two groups would end up in similar places, then they might interact quite frequently.
}

{
Formally, let $\mathcal{A}$ be the adjacency matrix of the 23 neighbourhoods in Vienna, normalised by dividing each row by the number of adjacent neighbours of the corresponding location. Each row of $\mathcal{A}$ therefore sums to one. The entry $\mathcal{A}{ij}$ is the probability that a person in location $i$ moves to location $j$ when they randomly choose an adjacent location. Let $\mu_X$ be the distribution of the population from country $X$ across neighbourhoods, expressed as a row vector. The distribution after one random move to an adjacent neighbourhood is then given by $\mu_X \mathcal{A}$. Finally, for any two distributions, say $X$ and $Y$, we measure how different they are using $\sum{i=1}^{23} (\mu_{iX} - \mu_{iY})^2$.
}

{
To test whether foreigners who experience high segregation show lower segregation after moving to an adjacent neighbourhood, we examine Syrians and Germans in Vienna, two groups with a large presence in the city but belonging to different clusters (Figure \ref{ViennaSpatial}). Results show that Syrians and Germans tend to live in different districts of Vienna and remain far apart even after moving to an adjacent district. The one-step distribution for Syrians and Germans remains substantial. For example, for many Germans, the city centre is in an adjacent neighbourhood; although few people live there, it is relatively accessible from where they reside. By contrast, relatively few Syrians reach the city centre even after a move to an adjacent neighbourhood.

\begin{figure*}[bt!]
\centering
\includegraphics[width=\textwidth]{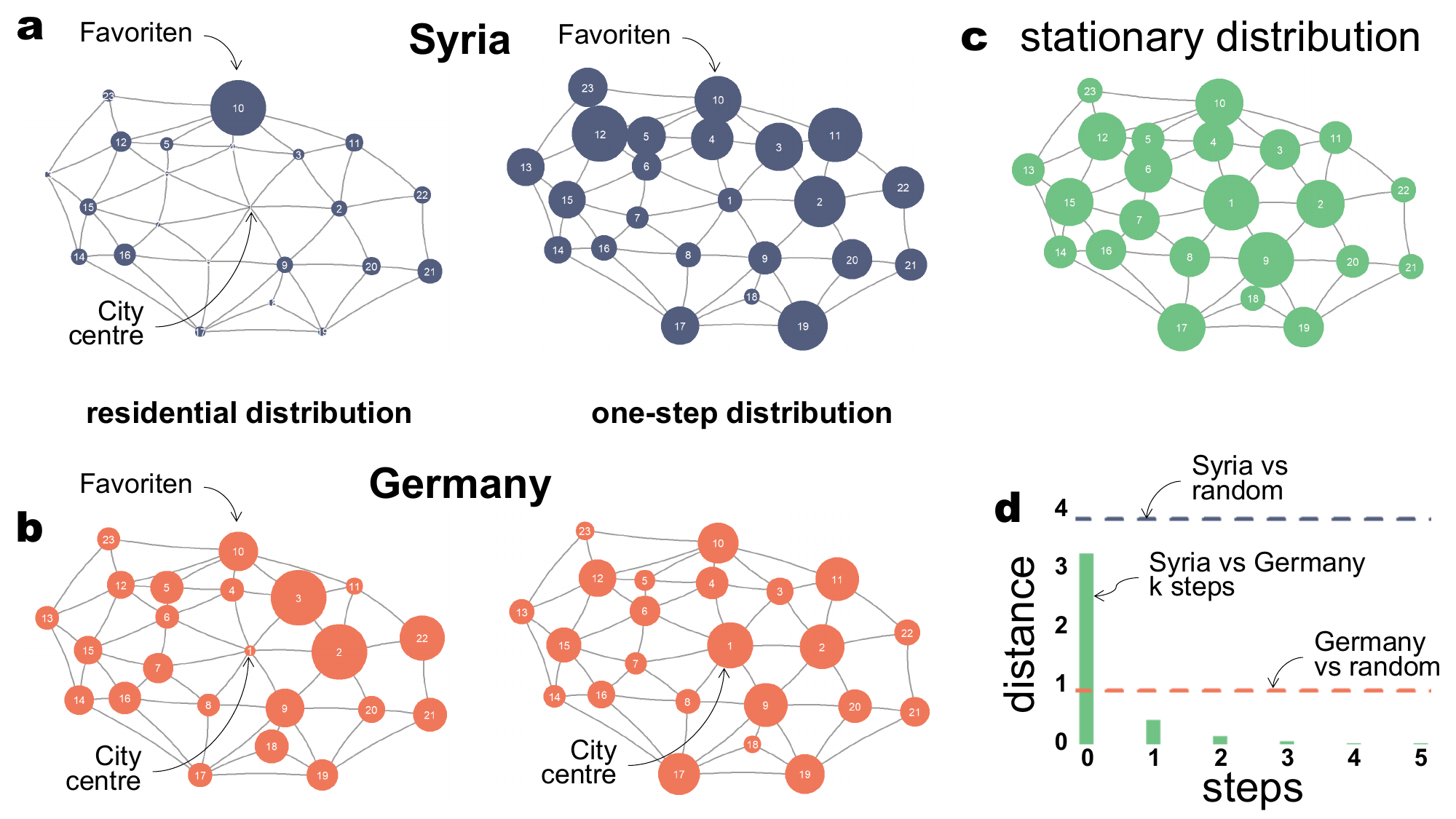}
\caption[Distribution of foreigners from Syria and Germany in Vienna.]{\textbf{Distribution of foreigners from Syria and Germany in Vienna.} \textbf{a} Residential and one-step distribution of people from Syria and \textbf{b} of Germany. Each node represents a district in Vienna, and two nodes are connected by an edge if they are adjacent. The size corresponds to the foreign population after moving to an adjacent district. \textbf{c} Stationary distribution, observed in the long-run, when people move each step to an adjacent neighbourhood randomly. \textbf{d} Distance between two distributions, measured by the combined difference across all districts. For reference, we added the difference between one nationality and a random distribution across the city.}
\label{ViennaSpatial}
\end{figure*}
}

{
Results show that the distribution of Germans in Vienna is much closer to a random distribution than that of Syrians. This suggests that Germans are more evenly spread across neighbourhoods, whereas Syrians are highly concentrated in a small number of neighbourhoods.
}

{
Since the adjacency matrix $\mathcal{A}$ is irreducible, meaning all districts are reachable through successive moves, and aperiodic, the process in which people move to an adjacent neighbourhood converges rapidly to its stationary distribution. This implies that, after a few steps, the distributions of different groups become increasingly similar and eventually cannot be distinguished within this framework. 
}

\section{Other measures of diversity and segregation} \label{sec:othermeasures}
Here, we compute the entropy index as a comparative measure for neighbourhood diversity, offering an alternative perspective to the Simpson index. Additionally, we employ the Gini index as a comparative measure of national homophily, juxtaposing it with the Dissimilarity index. By evaluating these alternative indices, we aim to validate and enrich our understanding of the patterns of diversity and segregation observed in Vienna.

\subsection{District diversity: Entropy index} \label{sec:entropy}
The information or uncertainty-based measure, otherwise known as the entropy index, is often referred to as the Shannon index, after the related work on information theory \cite{shannon1948mathematical, white1986segregation}. The Entropy index, which can be calculated for each district (local) and for the entire city, is a measure of concentration. The Entropy index $H^d$ for a district $d$ is \cite{white1986segregation}:
\begin{equation}
H^d=-\sum_{i} P_{i}^d \log \left(P_{i}^d\right),
\label{eq:entropy}
\end{equation}
where $P_i^d=N_i^d/N^d$ is the fraction of inhabitants of nationality \textit{i} residing in district \textit{d}. It is a measure between 0 and a maximum value, which depends on the number of different nationalities, $C^d$, residing in the district. The higher $H^d$, the more nationality-diverse the district is. A district is fully diverse ($H^d_{\text{max}}=\log(C^d)$) when the population in the district $d$ is equally distributed among all existing nationalities. In contrast, a district has no diversity ($H^d_{\text{min}}=0$), when it is inhabited exclusively by a single nationality.

Moreover, the city-wide Entropy index can be derived analogously to Eq. (\ref{eq:entropy}) by using the full population distribution in Vienna. The value for the city of Vienna is $H_\text{City}$ = 1.518. This number serves as a reference for comparing with the average local Entropy index, which is the population-weighted average of the local entropy index values across all districts. It is computed as \cite{white1986segregation}:
\begin{equation}
\Bar{H}=\sum_d\left(\frac{N^d}{N_{\text {City}}} H^d\right).
\label{eq:Averageentropy}
\end{equation}

When the average local Entropy index is substantially lower than the city-wide Entropy index, it indicates a high degree of segregation, with districts tending to be highly homogeneous, often dominated by a single nationality. In Vienna, the average local Entropy index is $\Bar{H} = 1.492$. This value closely aligns with the city as a whole, suggesting that segregation levels in Vienna are relatively low and that districts exhibit similar compositions. These results are presented in Supplementary Fig. \ref{fig:figs12}\textbf{a}, where districts are ranked in descending order of diversity. The Entropy index is highest for district 15 and lowest for district 13 (Supplementary Table \ref{tab:tabs3} for detailed values). Additionally, the colour bar highlights the difference between each district's diversity and the city-wide Entropy index, providing a clear comparison of which districts experience greater or lesser diversity relative to the city as a whole (Supplementary Fig. \ref{fig:figs12}\textbf{b}).

There is a strong correlation between the Simpson and Entropy indices, indicating that the Entropy index provides results consistent with the Simpson index (Fig. \ref{fig:fig4}\textbf{a}), and effectively captures the same qualitative patterns of district-level diversity. This correlation underscores the robustness of our findings, confirming that the observed patterns of district-level diversity remain consistent across different metrics.

\subsection{National homophily: Gini index} 
The Gini index, much like the Dissimilarity index, focuses solely on absolute inequalities by evaluating the concentration of a given nationality within specific districts, without accounting for the geographic proximity or spatial relationships between districts. In essence, the Gini index quantifies the degree of segregation experienced by each nationality across Vienna’s districts. It measures whether members of a particular nationality tend to reside predominantly in a small number of districts or are more evenly distributed throughout the city. The Gini index shares a strong conceptual connection with the Dissimilarity index. It represents the area between the Lorenz curve and the diagonal line of perfect equality, expressed as a proportion of the total area beneath the diagonal. This geometric relationship highlights the similarity between the two measures in their ability to capture inequality \cite{white1986segregation}.

The Gini index accounts for all pairwise comparisons in the system. For each nationality \textit{i}, $G_i$ can be computed as \cite{white1986segregation}:
\begin{equation}
G_i=\frac{1}{2 N_{City}^2 P_i\left(1-P_i\right)} \sum_{d} \sum_{d'} N^d N^{d'} \left|P_{i}^d-P_{i}^{d'}\right|,
\label{eq:gini}
\end{equation}
where $P_i=N_i/N_{City}$ is the fraction of inhabitants of nationality \textit{i} residing in Vienna city.

\begin{figure*}[bt!]
\centering
\includegraphics[width=\textwidth]{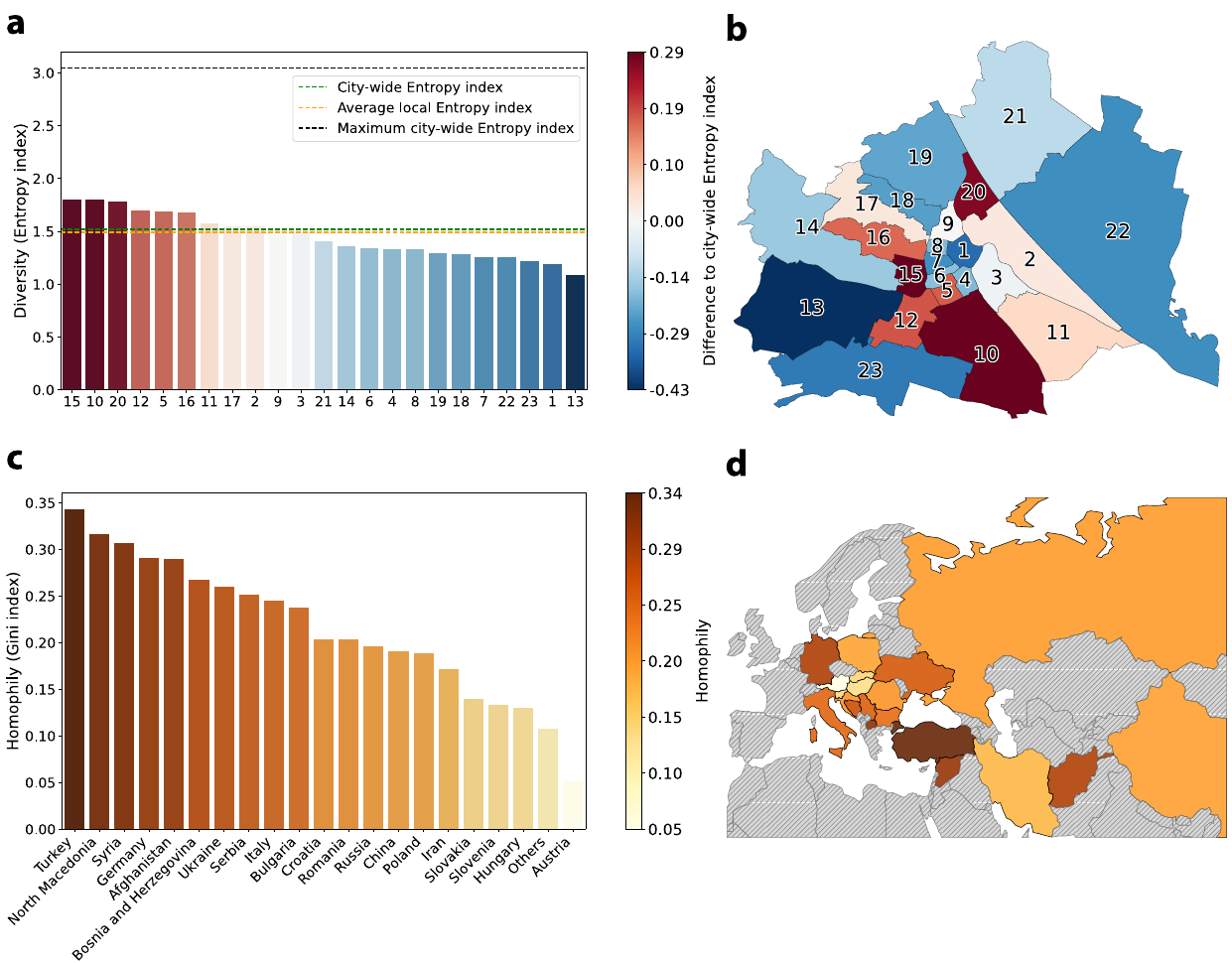}
\caption[District diversity and national homophily in Vienna using Entropy and Gini indices.]{\textbf{District diversity and national homophily in Vienna using Entropy and Gini indices.} \textbf{a} District diversity as measured by the Entropy index. Districts are ranked from highest to lowest by estimated diversity in Vienna. \textbf{b} Vienna diversity map. \textbf{c} National homophily assessed using the Gini index. Nationalities are ranked from highest to lowest according to the estimated homophily among countries. \textbf{d} World homophily map.}
\label{fig:figs12}
\end{figure*}

This index ranges from 0 to 1. A nationality is fully segregated ($G_i = 1$) when all the inhabitants of that origin reside in the same district. Conversely, a nationality is completely unsegregated ($G_i = 0$) when its population is evenly distributed across all districts (Supplementary Fig. \ref{fig:figs12}\textbf{c}). The Gini index is highest for Turkish nationals and lowest for Austrians (Supplementary Table \ref{tab:tabs4}). The Gini values for each country qualitatively align with the Dissimilarity indices, showing a strong correlation. This further corroborates the findings on national homophily by employing alternative methodologies.

Turkey exhibits the highest Gini index (Supplementary Fig. \ref{fig:figs4}). This is not only due to the significant disparities in residence concentration across districts, as shown on the map, but also to the pronounced absolute difference between the highest and lowest residence concentrations, as indicated by the colour bar. In contrast, Hungary exhibits the opposite pattern, with a much more uniform population distribution across districts. Similar patterns emerge for diversity and homophily when the Entropy and Gini indices are used as alternatives to our primary metrics. This demonstrates that our results are robust across different methods of measuring district diversity and national segregation. However, we opted to use our primary metrics rather than Entropy and Gini because of their simplicity, which aids visualisation and modelling.

Additionally, other indices, such as Moran's I, could be explored to examine neighbourhood segregation from a different perspective. Specifically, this approach would allow us to investigate how the spatial arrangement of districts and the population concentration in neighbouring districts influence national homophily and residents' decisions on where to live. By incorporating spatial autocorrelation, we can gain deeper insights into how proximity and district-level relationships contribute to patterns of segregation and residential choices.

\clearpage

\bibliographystyle{naturemag}
\bibliography{segregation}